\documentclass[preprint2]{aastex}
\usepackage{amssymb, amsmath, epsf, graphicx,multicol}
\usepackage{color}
\usepackage{graphicx}
\RequirePackage{lineno}

\def\vel{\rm \,km\,s^{-1}}
\def\be{\begin{equation}}
\def\ee{\end{equation}}
\catcode`\@=11 
\def\@versim#1#2{\vcenter{\offinterlineskip\ialign{$\m@th#1\hfil##\hfil$\crcr#2\crcr\sim\crcr } }}
\def\lsim{\mathrel{\mathpalette\@versim<}}
\def\gsim{\mathrel{\mathpalette\@versim>}}

\newcommand{\zhy}[1]{{ #1}}

\begin{document}
\title{Unshifted Metastable \zhy{He~I* Mini-Broad} Absorption Line System \zhy{in} the Narrow Line Type 1 Quasar SDSS J080248.18$+$551328.9}
\author{Tuo Ji\altaffilmark{1,2,3},
Hongyan Zhou\altaffilmark{1,2,3},  Peng
Jiang\altaffilmark{2,3,1}, Tinggui Wang\altaffilmark{2,3,1}, Jian
Ge\altaffilmark{4}, Huiyuan~Wang\altaffilmark{2,3,1}, S.
Komossa\altaffilmark{5,6,7}, Fred Hamann\altaffilmark{4}, Jens
Zuther\altaffilmark{8}, Wenjuan Liu\altaffilmark{1,2,3},
Honglin~Lu\altaffilmark{2,3}, Wenwen Zuo\altaffilmark{9}, Chenwei Yang\altaffilmark{1,2,3}, and Weimin~Yuan\altaffilmark{5,10}}

\altaffiltext{1}{Polar Research Institute of China, 451 Jinqiao Road,
Pudong, Shanghai 200136, China, jituo@pric.gov.cn}
\altaffiltext{2}{Key Laboratory for Research in Galaxies and Cosmology, The University of Science and Technology of China, Chinese Academy of Sciences, Hefei, Anhui, 230026, China}
\altaffiltext{3}{Center for Astrophysics, University of Science and Technology of China, Hefei, Anhui, 230026, China}
\altaffiltext{4}{Department of Astronomy, University of Florida,
Gainesville, FL 3261, US} \altaffiltext{5}{National Astronomical
Observatories, Chinese Academy of Sciences, 20A Datun Road, Beijing,
100012, China} \altaffiltext{6}{Excellence Cluster Universe,
Technische Universitaet Muenchen, Boltzmannstrasse 2, 85748
Garching, Germany} \altaffiltext{7}{Max-Planck-Institut fuer Radioastronomie,
Auf dem Huegel 69, 53121 Bonn, Germany} \altaffiltext{8}{Physikalisches Institut,
Universit\"at zu K\"oln, Z\"ulpicher Strasse 77, 50937 K\"oln,
Germany} \altaffiltext{9}{Department
of Astronomy, Peking University, Beijing 100871, China}
\altaffiltext{10}{National Astronomical Observatories/Yunnan
Observatory and Key Laboratory of the Structure and Evolution of
Celestial Objects, Chinese Academy of Sciences, Kunming 650011,
China}
\begin{abstract}

We report the identification of an unusual absorption line system
\zhy{in} the quasar SDSS J080248.18$+$551328.9 and present a
detailed study of the system, incorporating follow-up optical and
NIR spectroscopy. A few tens of absorption lines are detected,
including He I*, Fe II* and Ni II* that arise from metastable or
excited levels, as well as resonant lines in Mg~I, Mg II, Fe II, Mn
II, and Ca II. All of the isolated absorption lines show the same
profile of width \zhy{$\Delta v\sim 1,500\vel$} centered at a common
redshift as that of the quasar emission lines, such as [O II], [S
II], and \zhy{hydrogen} Paschen and Balmer series. With narrow
Balmer lines, strong optical Fe II multiplets, and weak [O III]
doublets, its emission line spectrum is typical for that of a
narrow-line Seyfert 1 galaxy (NLS1). We have derived reliable
measurements of the gas-phase column densities of the \zhy{absorbing} ions/levels. \zhy{Photoionization modeling} indicates that the absorber has
a density of \zhy{$n_{\rm H} \sim (1.0-2.5)\times 10^5~ {\rm cm}^{-3}$}
and a column density of \zhy{$N_{\rm H} \sim (1.0-3.2)\times 10^{21}
~ {\rm cm}^{-2}$}, and is located at \zhy{R$\sim100-250$} pc from the
central super-massive black hole. The location of the absorber, the
symmetric profile of the absorption lines, and the coincidence of
the absorption and emission line centroid jointly suggest that the
absorption gas is originated from the host galaxy and is \zhy{plausibly}  accelerated
by stellar processes, such as stellar winds \zhy{and/or} supernova
explosions. The implications for the detection of such a peculiar
absorption line system in an NLS1 are discussed in the context of
co-evolution between super-massive black hole growth and host galaxy
build-up.

\end{abstract}

\newpage
\section{Introduction}
It is now generally believed that active galactic nuclei (AGNs),
including their high-luminosity analog quasars, are powered by a
supper-massive black hole (SMBH) fed by accretion flows. To
enable the fuel feeding process, the angular momentum of the
interstellar gas in the \zhy{central region of} host galaxy must be
largely removed to bring \zhy{the gaseous fuel}
 \zhy{into} the nuclear region. None-axisymmetric perturbation of
gravitational potential, such as that due to stellar bars and
interaction with companion galaxies, are proven to be efficient ways
to drive large amounts of gas inwards on spatial scale of kpcs and
at timescales of Gyrs \citep{2004cbhg.symp..186W,2009pgn..confE...9D}.
\zhy{However, molecular} mappings of nearby Seyferts (e.g. NGC
1068, \citealt{2009ApJ...691..749M}; NGC 1097, \citealt{2009ApJ...702..114D}), show that the gas inflow is
terminated at spatial scales of a few to tens of pcs. \zhy{Other mechanisms, rather than gravity, are
necessary to drive the circumnuclear inflows. Theoretically,} the stalled
\zhy{molecular} gas will \zhy{definitely result} in nuclear starbursts \citep{2009ApJ...696..448H}.
\zhy{The energetic stellar processes of circumnuclear starburst might funnel gas further inwards to intra-pc scale
to fuel the black hole accretion \citep{2007ApJ...671.1388D,2009MNRAS.393..759S}.}

The fueling  \zhy{processes} should manifest themselves
more \zhy{observable} in AGNs with high mass accretion rate. A sub-class of
AGNs, namely narrow-line Seyfert 1 galaxies
(NLS1s)\footnote{\zhy{Following \cite{2006AJ....132..531K}, we collectively
speak of NLS1s when referring to the class properties of narrow line
Seyfert 1 galaxies and narrow line type 1 quasars; and we
distinguish between a Seyfert galaxy and a quasar according to the
classical criterion of B-band absolute magnitude when referring to
individual objects.}}, are  \zhy{
generally considered to be AGNs at their early evolutionary stage with small
black hole masses accreting at very close to the maximum allowed accretion rate
(see \citealt{2008RMxAC..32...86K} for an extensive review)}. \zhy{They} are
traditionally defined by the narrowness of their Balmer emission
lines (FWHM$_{{\rm H}\beta}<$2000 km s$^{-1}$) and weakness of [O
III] emission ([O~III]/H$\beta <3$; \citealt{1985ApJ...297..166O},
but see also \citealt{2006ApJS..166..128Z} for a slightly different
definition). Strong Fe II emission is also \zhy{a significant feature} for NLS1s.

\zhy{\cite{2000MNRAS.314L..17M} suggested that NLS1s live in gas-rich
  galaxies with ongoing star formation. Observationally, star formation in the host galaxies of NLS1s is considerable
stronger compared with that of the normal type 1 AGNs
 using Spitzer mid-infrared spectroscopy \citep{2010MNRAS.403.1246S} . }
The observations can be merged into the scenario of co-evolution of galaxy and
central black hole \cite[e.g.][]{2004ApJ...600..580G}. The energetic stellar processes
of circumnuclear starburst might induce the surrounding gas flowing inwards
more efficiently to trigger high activity of the central black hole \citep{2007ApJ...671.1388D,2009MNRAS.393..759S}.
\zhy{Provided that the gaseous inflows would occasionally intercept our line of sight
toward active nucleus, they can be observable through quasar absorption line technique.}

 In this paper, we report the first such candidate absorption line system
toward the \zhy{narrow line type 1} quasar SDSS
J080248.18$+$551328.9 (hereafter SDSS~J0802$+$5513 for brevity),
 and present detailed study of the
system incorporating follow-up optical and NIR spectroscopy.

\zhy{SDSS~J0802$+$5513 was initially identified as a quasar during
the Sloan Digital Sky Survey (SDSS, \citealt{2000AJ....120.1579Y})
based on spectroscopic observation on 2003-03-24, and was included
in the SDSS quasar catalogue with a redshift of $z=0.6640\pm0.0005$
\citep{2007AJ....134..102S, 2010AJ....139.2360S}. It was first
classified as a low-ionization broad absorption line (LoBAL) quasar
by \citet{2009ApJ...692..758G} according to detection of Mg~II broad
absorption trough with width $\Delta v \approx 2,500$~km~s$^{-1}$. The
authors also reported possible detection of
Fe~II $\lambda\lambda$2414, 2632, 2750 absorption lines.
SDSS~J0802$+$5513 was later analyzed by \citet{2010ApJ...714..367Z}
but rejected as a LoBAL quasar since the Mg~II absorption line does
not show a large enough blueshift, though the absorption trough is
broader than their threshold value of $\Delta v > 1,600\vel$. Using
Kohonen self-organizing maps, \citet{2012A&A...541A..77M} again
classified SDSS~J0802$+$5513 as an unusual LoBAL quasar with a red
color and narrow absorption line troughs, and confirmed the
classification by visual inspection of its SDSS spectrum. The
discrepant classification of the absorption lines in the quasar
deserves further exploration.}

\zhy{We noticed SDSS~J0802$+$5513 during our systematic search for
He~I* $\lambda \lambda$2765, 2830, 2946, 3189, 3889, 10830 absorption
multiplets in quasars\footnote{In this paper, we will use `*' to
denote absorption lines that arise from metastable/excited levels.
E.g., metastable neutral helium absorption lines will be referred to
as He~I*, and singly-ionized iron absorption lines from excited
levels as Fe~II*.} \citep[][submitted to ApJS]{liu2014}.
Arising from the common metastable He I* 2 $^3$S level, the multiplets are
rarely seen in the interstellar medium of normal galaxies (Rudy et
al. 1985). The 2 $^3$S level is mainly populated by recombination of
He$^+$. With an ionization potential of 4.8 eV, the diffuse stellar
radiation background can easily ionize helium atoms at the 2 $^3$S
level, while  lacking of enough hard photons of $h\nu
>24.59$ eV to ionize helium atoms at the ground level.
Whereas quasar continuum is energetic enough to populate a large
number of He atoms at the 2 $^3$S level. He~I* multiplets are hence
a good indicator for distinguishing quasar intrinsic narrow
absorption lines (NAL) from intervening NALs that are physically
unrelated to the background quasars, in complementary to the two
often used indicators \citep[e.g.][]{1997ApJ...478...87H,2003AJ....125.1336M}: (1) time variability and (2) partially coverage
of the absorption lines. He~I* multiplets are very useful for
determining the covering factor of quasar absorption gas due to the
large oscillator strength differences  in the
multiplets. This, together with their common highly metastable lower
transition 2 $^3$S level\footnote{\zhy{The asymptotic maximum value of
$n_{2~^3{\rm S}}/n_{{\rm He}^+}\approx 10^6$ for a temperature of $10^4$ K and
an electron density of $n_{\rm e}\gtrsim 10^4$~cm$^{-3}$ when collisional
de-excitation rate dominates radiative rate (Rudy 1985).}}, endue
He~I* multiplets with a high sensitivity to a large dynamic range of column
densities when other strong UV resonant lines, such as O~VI, N~V,
C~IV, Si~IV, Al~III and Mg~II, are heavily saturated \citep{2001ApJ...561..118A,2011ApJ...728...94L,2014ApJ...788..123L}. The wide wavelength separations of
the lines in He~I* multiplets also relieve line blending, which is
always a severe problem for BAL studies. Clear detection of
He~I* $\lambda\lambda$3189, 3889, 10830 and a few tens of other
absorption lines, including Ni~II*, Fe~II*, Fe~II, Mn~II, Ca~II,
Mg~II and Mg~I, enable us to derive reliable measurements of the
column densities of the corresponding ions/levels; to probe the
physical conditions of the absorption gas; and to locate the gas
with the aid of photoionization model calculations.}

The paper is organized as follows. In \S2, \zhy{we will present a
brief description of the SDSS quasar sample with He~I* absorption
line multiplets, and identify SDSS~J0802$+$5513 as an outlier in the
sample}. We will also describe there the data used in this paper,
including collection of archived data and observations and data
reductions of our new spectroscopies. Measurements of emission and
absorption lines will be presented in \S3. In \S4, we will perform
extensive photoionization simulations with the spectral synthesis
code CLOUDY \citep{1998PASP..110..761F}, and make comparison between
model calculations and absorption line measurements. Possible origin
of the absorption gas and its implications will be discussed in \S5.
Our main results will be summarized in the last section, together
with future perspectives. Throughout this paper, we assume a
cosmology with $\Omega_\Lambda$= 0.7, $\Omega_M$ = 0.3, and
H$_{0}$=70~km~s$^{-1}$~Mpc$^{-1}$. \zhy{All  quoted magnitudes are
using AB-system except 2MASS ones, which are Vega magnitudes. We use
negative velocities to denote blueshifted absorption lines, and
positive values for redshifted lines.}

\section{\zhy{Identification, }Observations and Data Reduction\label{sec:data}}

\subsection{\zhy{SDSS J0802$+$5513 as an Outlier of Quasar He~I* Absorbers}}
\begin{figure*}[!ht]
\hspace*{1cm}\vspace*{0cm}\includegraphics[angle=0,scale=1]{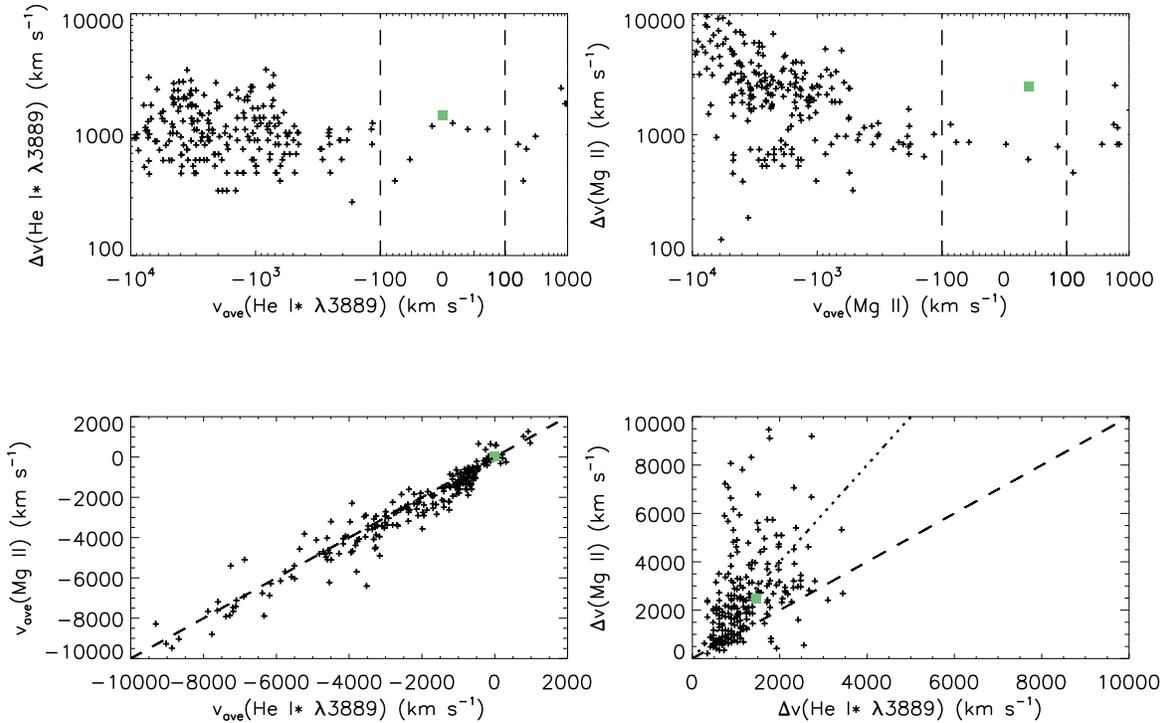}
\caption{\zhy{Upper left: He I* $\lambda3889$ absorption line velocity
    shifts $v_{\rm ave}$ versus their widths $\Delta v$ for quasar He I* absorbers identified from SDSS DR7. For absorbers with $|v|>$100,  abscissa  is in log scale, while for absorbers with $|v_{\rm ave}|<$100 in linear scale. SDSS
J0802+5513 is  shown as a green square. Two vertical dashed lines
at $\pm$100 km s$^{-1}$ are shown to guide the eyes. Absorbers in
between the two vertical lines are deemed as unshifted, for the dispersion is in agreement with uncertainties of systematic redshift determination for SDSS quasars. Upper right: same as upper left except for Mg II lines. The velocity and width of  the two lines are compared in the lower two panels (lower left: velocity. lower right: width). Note that the velocities of the two lines almost have the same values as indicated by the dashed line, while the widths of Mg II are much larger than that of He I (Mg II:He I=1:1 and 1:2 are shown as dashed and dotted lines, respectively, to guide the eyes).} \label{fig:v_dv}}
\end{figure*}

\zhy{He~I* multiplets are much under-explored as an important
diagnostics for the ionization state of quasar absorption gas. So
far only a handful detections have been reported in the literature
(see \citealt{liu2014} for a summarization). He~I*~$\lambda$10830,
located in the NIR and the strongest line in the multiplets, is beyond the
wavelength coverage of most large sky area optical spectroscopic
surveys. With a $\lambda f_{ik}$ value of 23.3 times less than
He~I*~$\lambda$10830 \citep{2011ApJ...728...94L}, the second strongest
line, He~I*$\lambda$3889, though falling in the optical for quasars
at $z\lesssim 1$, is much weaker than those familiar UV resonant
lines. In that wavelength range, the pseudo-continuum from complex
iron multiplets, which usually varies a lot from object to object,
makes the normalization quite uncertain. Detection of such a weak line with contamination
is  a challenging task. We developed a new detection technique  to minimize the uncertainty of normalization.
We use the observed quasar spectra in a selected
library to mimic the absorption free spectra of the quasar of interest,
in analog to the pair-method traditionally used to
recover the extinction-free stellar spectra (see \citealt{liu2014} for
a full description of the method). The `pair-method' can
yield a much better precision than the often-adopted approaches for
quasar absorption line measurements. Using this newly developed
technique, we detected He~I*~$\lambda$3889 line at $> 2~\sigma$ confidence level in
247 quasars with Mg~II absorption from the Seventh Data
Release (DR7) of SDSS. The quoted uncertainties include both 
statistical and systematic errors. As shown in Figure 1, the
velocity centroid of Mg~II and He~I* $\lambda$3889 agrees with each
other within 10\%. This indicates that the vast majority of
absorption lines in the sample are real detections, since Mg~II and
He~I* $\lambda$3889 are measured independently. In this Figure, we
also compared the width of Mg~II and He~I* $\lambda$3889, and found
that the former is on average 2 times broader than the later with a
rather large scatter. This fact was also noted by \citet{2001ApJ...561..118A}
in a case study of QSO 2359$-$1241. }

\zhy{The sample show interesting pattern in the Mg~II line velocity
against width ($v_{\rm ave}$-$\Delta v$) \footnote{\zhy{$v_{\rm ave}$ is the
    absorption-weighted average velocity, while $\Delta v$ is the
    difference between maximum and minimum velocities of the
    absorption troughs, See Liu et al. 2014 for details.}} plane of Figure 1. Line width is
strongly correlated with velocity for absorbers in the upper-left
corner ($\Delta v$ and $v_{\rm ave}$ $\gtrsim$ 1,000~km~s$^{-1}$), which can be
readily classified as BALs. Absorbers located in the lower-right
corner should fall in the traditional intrinsic NAL category.
Considering the velocity zero-point uncertainty of $\sim$
100~km~s$^{-1}$ determined by [O~III], [O~II], Mg~II, and Balmer
emission lines \citep{2006ApJS..166..128Z,2008MNRAS.386.2055N},
absorbers located in the middle part within the two dashed lines can
be taken as unshifted, most of which might be of a different
origin than the left-hand BAL and NAL outflows. Their relation with
those redshifted absorbers located in the lower-right corner is not
clear, neither is the nature of the redshifted absorbers (see \citealt{shi2014} for a case study)\footnote{\zhy{The origin of these
redshifted absorbers may also be different from the BAL quasars with
redshifted troughs identified by \citet{2013MNRAS.434..222H}, which are much
broader and often with blueshifted absorption too.}}. With an
associated quasar showing an NLS1-like emission line spectrum, the
broadest Mg~II absorption trough, a common redshift coincidence with
that of the quasar emission lines, the same symmetrical velocity
structure of all isolated absorption lines, SDSS J0802$+$5513 is the
most extreme case of the unshifted He~I* $\lambda$3889 absorbers.
Taking advantage of the profuse absorption line spectrum, a detailed
study of this extreme case may shed new light on the nature of this
mysterious new kind of absorbers. We collected existing photometric
data and performed subsequent NUV through optical to NIR spectroscopic follow-ups from
2008 to 2012 using the Blue Channel Spectrographs at the Multiple
Mirror Telescope (MMT), the Multi-Object Double Spectrographs (MODS)
at the Large Binocular Telescope Large Binocular Telescope (LBT),
and TripleSpec at the 200-inch Hale Telescope (P200). The
photometric and spectroscopic data are summarized in Table
\ref{tbl:photometry} and Table \ref{tbl:spectroscopy},
respectively. We show the data in Figure \ref{tbl:spectroscopy} and give them  detailed descriptions in next two subsections.}

\subsection{Broad Band Photometry\label{sec:photometry}}
\begin{figure*}[!ht]
\vspace*{0cm}\includegraphics[angle=0,scale=0.8]{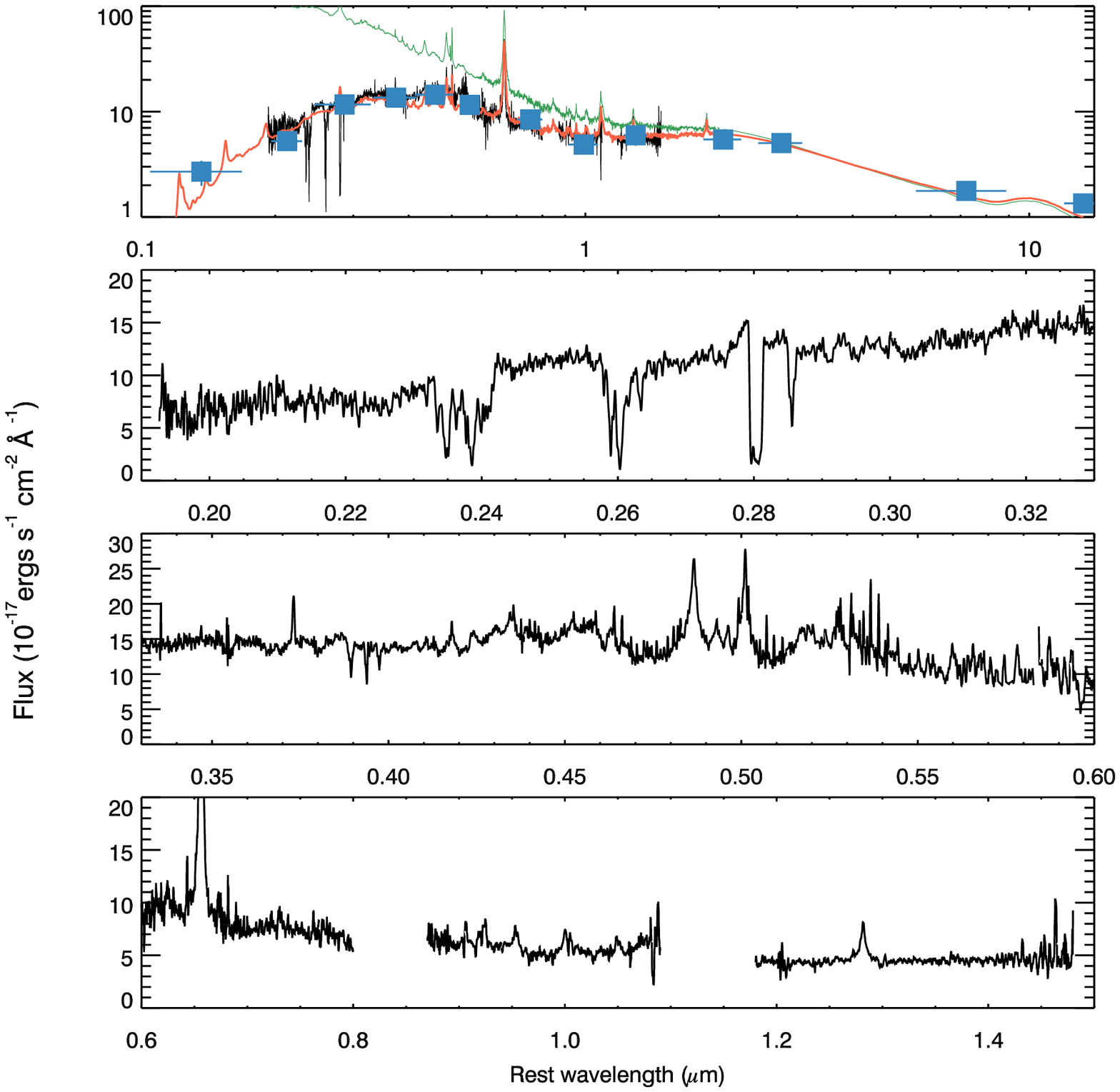}
\caption{\zhy{In first panel, merged spectrum of SDSS J0802+5513 in the quasar rest-frame
is shown in black curve. The data used include SDSS, MMT, LBT and
P200 spectroscopy (see Table \ref{tbl:spectroscopy} and
\S~\ref{sub:spec} for detail), which are smoothed by a boxcar of 7
pixel with bad pixels removed for clarity. A quasar composite
spectrum is  overplotted in green for comparison. Broad band spectral energy distribution of SDSS
J0802+5513 (black squares, photometric data are adopted from GALEX,
SDSS, 2MASS and WISE; see Table \ref{tbl:photometry} for detail) is
displayed  along with the
quasar composite reddened by $E(B-V)=0.36$ using the SMC-type
extinction law (red curve). Expanded views of the merged spectrum are shown in subsequent panels.}\label{fig:sed}}
\end{figure*}

The photometric data include fluxes at 1.4 GHz by NRAO VLA
Sky Survey (NVSS, \citealt{1998AJ....115.1693C}) and Faint Images of the Radio
Sky at Twenty-cm (FIRST, \citealt{1995ApJ...450..559B}) in the radio; IR
magnitudes from the Wide-field Infrared Survey Explorer (WISE; \citealt{2010AJ....140.1868W}) and the Two Micron All Sky Survey (2MASS,
\citealt{2006AJ....131.1163S}); optical magnitudes from SDSS; and NUV
magnitude from the Galaxy Evolution Explorer (GALEX, \citealt{2007ApJS..173..682M}).

 \zhy{SDSS J0802$+$5513 has been monitored in Catalina Real-time Transient
Survey over 7 year \citep{2012IAUS..285..306D}. The light curve in V band shows no
significant variations beyond the intrinsic 1-$\sigma$ flux scatter of 15\% (see Figure 3). The data suggests that the quasar did not show
significant variability. As shown in the first
  panel of Figure \ref{fig:sed}, the SED of SDSS J0802+5513 is
  apparently red as compared to a quasar composite. The quasar composite used in the fit is the
combination of SDSS optical ($\lambda\leq3000$
\AA, \citealt{2001AJ....122..549V}), IRTF NIR (3000
\AA$\leq\lambda\leq3.5$ $\mu$m, \citealt{2006ApJ...640..579G}; see
\citealt{2010ApJ...708..742Z} for an application) and Spitzer MIR to FIR ($\lambda\ge
3.5$ $\mu$m, \citealt{2007ApJ...666..806N}) composites. We then  estimated reddening of
the quasar by fitting the quasar composite spectrum to the SED of SDSS J0802$+$5513 assuming a Small
Magellanic Cloud (SMC) type extinction curve \citep{1982A&A...113L..15L,2005ApJ...630..355C}}. The best-fitted model
is obtained by minimizing $\chi^2$.  The best-fitted $E(B-V)$ is 0.36
mag, which agrees well with that indicated by the flux ratios of
Pa$\beta$/H$\alpha$ (E(B$-$V)$\sim$0.4) and H$\alpha$/H$\beta$ (E(B$-$V)$\sim$0.3) as detailed in \S3.1. It
can be seen \zhy{in the first panel of} Figure \ref{fig:sed} that this simple model
well reproduces the observed SED.

SDSS J0802+5513 is detected by FIRST and NVSS at 1.4 GHz in the
radio. No significant variability \zhy{was observed} between the two \zhy{ epochs}.
{Radio
loudness, defined as $R\equiv S_{\rm 5\ GHz}/S_B$, is 110 and 25,
 respectively, before
and after reddening correction. The NVSS flux at 1.4 GHz was used to
estimate the 5 GHz emission $S_{\rm 5\ GHz}$ and a radio spectral index
of $\alpha_{\rm r} = 0.5$ ($S_{\nu}\propto \nu^{-\alpha_{\rm r}}$) was assumed.
We corrected the $B$ band flux $S_B$ flux using the best-fitted
reddening value. SDSS J0802+5513 is moderately radio-loud, and it
becomes radio-intermediate after reddening correction.} SDSS
J0802$+$5513 was also observed at 18 cm by the
Multi-Element-Radio-Linked Interferometer Network (MERLIN). It is
unresolved at $0''.3$ resolution, corresponding to a linear scale of
about 1.6 kpc the redshift of the quasar \citep{2012JPhCS.372a2005Z}.

\subsection{Spectroscopy\label{sub:spec}}

\begin{deluxetable}{ccclc}
\tabletypesize{\footnotesize}
\tablecaption{Broad Band Photometry of SDSS J0802+5513 \label{tbl:photometry}}
\tablewidth{0pt}
\tablehead{
\colhead{band} & \colhead{flux} & \colhead{facility} & \colhead{obs. date} &\colhead{reference}\\
\colhead{ } & \colhead{mag/mJy} & \colhead{ } & \colhead{(UT)} &\colhead{ } }
\startdata
NUV & 22.62$\pm$0.26 & GALEX & 2004-06-05 & 1\\
$u$ & 20.79$\pm$0.06 & SDSS &  2003-03-11 & 2 \\
$g$ & 19.21$\pm$0.02 & SDSS &  2003-03-11 & 2 \\
$r$ & 18.42$\pm$0.01 & SDSS &  2003-03-11 & 2 \\
$i$ & 17.88$\pm$0.02 & SDSS &  2003-03-11 & 2 \\
$z$ & 17.70$\pm$0.02 & SDSS &  2003-03-11 & 2 \\
$J$ & 16.47$\pm$0.10 & 2MASS & 1999-04-27 & 3 \\
$H$ & 15.94$\pm$0.13 & 2MASS & 1999-04-27 & 3 \\
$K_s$ & 14.66$\pm$0.09 & 2MASS & 1999-04-27 & 3 \\
$w_1$&12.93$\pm$0.03&WISE&2010-01-10&4\\
$w_2$&11.72$\pm$0.02&WISE&2010-01-10&4\\
$w_3$&8.84$\pm$0.03&WISE&2010-01-10&4\\
$w_4$&6.46$\pm$0.04&WISE&2010-01-10&4\\
$1.4$ GHz&7.1$\pm$0.5 &NVSS&1993-11-23&5\\
$1.4$ GHz&6.54$\pm$0.14 &FIRST&1998-07-17&6
\enddata
\tablerefs{(1) \citealt{2007ApJS..173..682M}; (2) \citealt{2010AJ....139.2360S}; (3)
\citealt{2006AJ....131.1163S}; (4) \citealt{2010AJ....140.1868W}; (5)
\citealt{1998AJ....115.1693C}; (6) \citealt{1995ApJ...450..559B}.}
\end{deluxetable}

\begin{deluxetable}{cccccc}
\tabletypesize{\small} \tablecaption{Journal of Spectroscopic
Observations \label{tbl:spectroscopy}} \tablewidth{0pt} \tablehead{
\colhead{wavelength coverage} & \colhead{slit} & \colhead{resolution}&\colhead{exposure}&\colhead{telescope/instrument}&\colhead{obs. date}\\
\colhead{(\AA)} & \colhead{(arcsec)} &
\colhead{$\lambda/\Delta\lambda$}&\colhead{(sec)}&\colhead{}&\colhead{UT}
} \startdata
3800--9200&3\tablenotemark{a}&2000&4200&SDSS 2.5m&2003-03-24\\
3200--5200&1&\zhy{1800}&1500&MMT/Blue Channel&2008-03-30\\
10000--24000&1.1&\zhy{2500}&1200&P200/TripleSpec&2011-10-21\\
3200--11000&0.8&\zhy{2500}&\zhy{2400}&LBT/MODS&2012-01-29\\
\enddata
\tablenotetext{a}{The SDSS spectrograph is fiber fed with a size
of $3^{''}$ in diameter. }
\end{deluxetable}

The SDSS spectrum that we use is the improved sky-residual subtracted version
as published in \citet{2010MNRAS.405.2302H}. 
\begin{figure}[!ht]
\vspace*{0cm}\includegraphics[angle=0,scale=0.5]{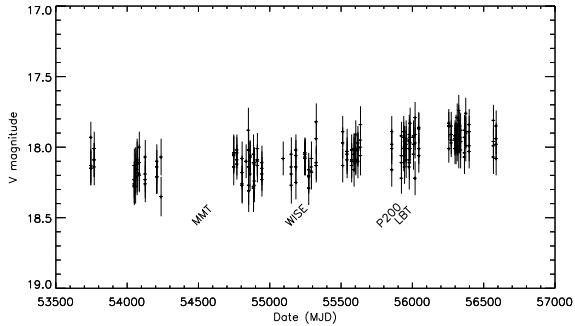}
\caption{\zhy{Catalina V-band light curve for SDSS J0802+5513.  Observing MJDs for all the data  are labelled by telescope names, with the exception of SDSS and GALEX, whose observing epochs are not covered by Catalina.} \label{fig:lightcurve}}
\end{figure}
{To detect any possible absorption line variations and achieve a
higher signal-to-noise ratio (S/N),} We carried out a follow-up NUV
spectroscopic observation at MMT using the blue channel on
2008-03-30. We used the $1''\times180''$ slit and 800 l/mm grating.
A total observation time of 1500 s  \zhy{was} equally split into two
exposures. The seeing during the observation is about $1.0''$.
He/Ne/Ar  \zhy{lamp} is used for wavelength calibration and the KPNO
standard star eg182 is observed for flux calibration. We use the
standard IRAF package\footnote{IRAF is distributed by the National
Optical Astronomy Observatory, which is operated by the Association
of Universities for Research in Astronomy, Inc., under cooperative
agreement with the National Science Foundation.} to extract the 1-D
spectrum. The extracted spectrum covers a wavelength range of
$\sim$3200--5200 \AA. The median S/N of the MMT spectrum is about 10
pixel$^{-1}$ with a resolution of $R\sim 1800$, similar to that of
SDSS ($R\sim 2000$). Compared to SDSS spectrum, the MMT spectrum shows no
significant variations in absorption lines, neither the velocity
profiles nor the maximum depths (see \S\ref{sec:absorption} for
detail).

To observe H$\alpha$ emission line and the expected He I*
$\lambda$10830 absorption line, we \zhy{acquired} a NIR spectrum using
TripleSpec at P200 telescope via China Telescope Access Program
(TAP). The observation was carried out on 2011-10-21 using \zhy{the standard slit of $1''\times30''$ in
A-B-B-A dithering mode. The total exposure time was 1200 s.} Using IDL-based Spextool software
\citep{2004PASP..116..362C}, the raw data are flat-field corrected,
telluric corrected \citep{2003PASP..115..389V}, wavelength (using
sky lines) and flux calibrated. The \zhy{reduced} spectrum has
\zhy{ a} wavelength coverage of $\sim 1.0$--2.4 $\mu$m and a resolution of
$R\sim 2500$.
 
To fill the gap between SDSS and P200 spectra and to detect the
possible Na I D absorption doublet, we carried out a follow-up
spectroscopy using \zhy{ MODS at} LBT with a slit width of $0''.8$. Four 600 s
exposures \zhy{were} acquired for each of the blue and red channels. CCD
reductions, including bias subtraction, flat-field correction were
accomplished using Python package ``modsCCDRed''. Subsequent reductions
are carried out using standard IRAF package. Ne/Hg/Ar/Xe/Kr lamps
were used for wavelength calibrations and the standard star Ferge 67
was observed for flux calibration. The extracted LBT spectrum covers
 \zhy{ a} wavelength range of $\sim 3200$--$10000$ \AA~and a resolution of
$R\sim 2500$. \zhy{Na I D absorption doublets were not significantly detected on the
moderate S/N LBT spectrum.}

All \zhy{above-mentioned} spectra were corrected for the Galactic extinction
of $E(B-V)= 0.05$ \citep{1998ApJ...500..525S} assuming an \zhy{average Galactic ($R_V=3.1$)}
extinction law, and transformed into the quasar rest-frame using the
redshift of $z=0.6640\pm0.0005$ as determined by [O~II], [S~II],
Balmer and Paschen emission lines, which is in consistent with that ($z=0.6641\pm0.0004$) given
in \citet{2010MNRAS.405.2302H}  within errors.
\zhy{Since the quasar SDSS J0802$+$5513 shows no significant variation in years (see \S\ref{sec:photometry}),
we then recalibrated all spectra to photometric data for correcting aperture and seeing effects. We then
combined all the spectra, weighted by their S/N, to construct a broad band spectral energy distribution (SED).}
We display the \zhy{ combined spectrum in the first panel of Figure \ref{fig:sed} with expanded views in subsequent panels.} It shows narrower Balmer lines and
stronger optical Fe II emission lines. [O~II] emission is very
strong relative high ionization forbidden lines such as [Ne III] and
[O III]. Quantitative emission line measurement will be presented in
\S\ref{sec:emission}. Besides emission lines, the absorption
spectrum of SDSS J0802$+$5513 is profuse from rest-frame NUV through
optical to NIR. The most prominent absorption features are the
strong He I* line at 10830 \AA\ and the blended troughs of Fe II UV1
and UV2 around 2600 \AA\  and 2400 \AA\, respectively. Identification and measurement of
the absorption lines will be described in \S\ref{sec:absorption}.

\section{Spectral Analysis}
\subsection{Emission Lines Measurement\label{sec:emission}}
\begin{figure}[!ht]
\includegraphics[scale=0.55]{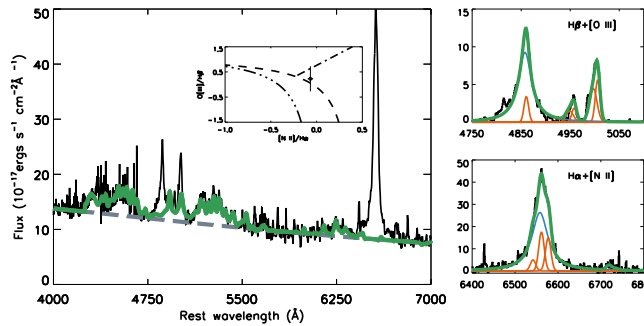}
\caption{Decomposition of the rest-frame optical spectrum of SDSS
J0802+5513. In the left panel, the observed pseudo-continuum (black)
is modeled by the combination (green) of a power law (dashed gray)
and Fe II template. The right panels display the emission line
spectrum in the H$\beta$+[O III] and H$\alpha$+[N II] regimes after
removal of the pseudo-continuum with fitting results overlaid. Narrow
forbidden lines are fitted by Gaussian(s), and H$\alpha$ and
H$\beta$ are fitted by the combination of a Lorentzian and a
Gaussian, representing the broad and narrow components,
respectively. Observed data are shown in black curves, narrow
emission lines in orange, broad lines in blue, and model
combinations in green (See \S3.1 for detailed description of the
models). Diagnostic diagram for narrow line measurements (black
diamond with error bars) are shown in the insert of the left panel.
Classification lines between LINERs and Seyfert 2 galaxies \citep{2006MNRAS.372..961K} is shown in dash-dot line, and division between AGNs
and star forming galaxies from \citet{2003MNRAS.346.1055K} and \citet{2006MNRAS.372..961K} in dashed and dotted lines, respectively.
\label{fig:emission}}
\end{figure}
\begin{deluxetable}{lcccc}
\tablecaption{ \zhy{Relevant} emission lines measurement\label{tbl:emission}}
\tablewidth{0pt}
\tablehead{
\colhead{Transition}&\colhead{Broad line flux}& \colhead{\zhy{Broad line Width}}&\colhead{Narrow line flux}&\colhead{\zhy{Narrow line Width}}\\
\colhead{} & \colhead{10$^{-17}$ ergs cm$^{-2}$~s$^{-1}$}& \colhead{\zhy{$\vel$}}&\colhead{10$^{-17}$ ergs cm$^{-2}$~s$^{-1}$}&\colhead{\zhy{$\vel$}}
}
\startdata
Pa$\beta$&230$\pm$20&1800$\pm170$&32$\pm$10&$690\pm120$\\
H$\alpha$&1946$\pm$28&1800$\pm170$&235$\pm$10&$690\pm120$\\
H$\beta$&487$\pm$15&1800$\pm170$&40$\pm$15&$690\pm120$\\
${\rm [O~II]}$&\nodata&\nodata &45$\pm$6&$690\pm120$\\
\enddata
\tablenotetext{a}{All quoted errors are statistical ones, widths are FWHMs, the widths of all broad emission lines and narrow lines are tied during the fit, respectively.}

\end{deluxetable}

We adopt a fitting method similar to \citet{2006ApJS..166..128Z}  and
\citet{2011ApJ...736...86D} to measure the emission lines of
interest, with some modifications to accommodate the existence of
broad absorption lines. Briefly, the observed spectrum was
decomposed into a power law, Fe II emission complex, and emission
lines other than Fe~II. \zhy{Given the SED of SDSS J0802+5513 is red, we fit a reddened power law  to emission line free windows: 4020--4050 \AA, 4205--4235 \AA, 5580--5620 \AA,
6350--6400 \AA, 6810-6850 \AA.} Then we subtract the best fitted
power law from the observed spectrum, and fit the residual by a Fe
II template. \zhy{ We employed the Fe II template} built by
\citet{2004A&A...417..515V}.  \zhy{ The template was broadened} to fit the Fe II emission
strong \zhy{ spectral} windows: 4170--4260 \AA, 4430--4770 \AA\, 5080--5500 \AA\ and
6050--6200 \AA. The best-fitted model of the pseudo-continuum,
including a power law and Fe II multiplets, is shown in the left
panel of Figure \ref{fig:emission}.

Emission lines other than Fe~II multiplets were measured by fitting
the pseudo-continuum-subtracted spectrum. Each of Balmer and Paschen
lines were modeled with two components, a  Lorentzian profile for the
broad line and a Gaussian for the narrow one. The narrow Balmer and
Paschen lines were assumed to have the same width and redshift as
that of the low-ionization forbidden lines [S~II] and [N~II]. All of
the narrow emission lines were fitted by a single Gaussian except
[O~III] doublet, which were fitted with two Gaussians, one for the
blue wing \citep{2008RMxAC..32...86K} and the other for the core of the
lines. The width and redshift of [O~III] core component were tied to
that of low-ionization narrow lines. The doublet ratios of [O~III]
and [N~II] are fixed to their theoretical value 3:1 during the fit.
\zhy{We zoomed in the best-fitted model in H$\alpha$ and H$\beta$ regimes in
the right panel of Figure \ref{fig:emission}.}

The measured line parameters are summarized in Table
\ref{tbl:emission}. \zhy{In  \citet{2008MNRAS.383..581D}, the authors
investigated the broad-line Balmer decrements  for an unreddened
sample of Seyfert 1 galaxies and QSOs in SDSS and  they find that  the
distribution of the intrinsic broad-line H$\alpha$/H$\beta$ ratio can be
well described by log-Gaussian, with the peak at H$\alpha$/H$\beta$=3.06 and a
standard deviation of about 0.03 dex only.}  The steep
broad line ratios of H$\alpha$/H$\beta=4$ \zhy{in SDSS J0802+5513} thus indicate that the broad line region (BLR) should
be significantly reddened. \zhy{Using
H$\alpha$/H$\beta$ as the intrinsic Balmer decrement and an SMC type
extinction curve, we obtain an
estimate of $E(B-V)=0.31\pm0.05$}.  \zhy{No estimation of intrinsic
Pa$\beta$/H$\alpha$ is available in \citep{2008MNRAS.383..581D}, we
assume  an intrinsic  Pa$\beta$/H$\alpha$  ratio of 0.06}
 \zhy{as calculated in case B} for typical BLR conditions (\zhy{i.e.,} $T_{\rm e}=10000$~K and $n_{\rm e}=10^{9}$~cm$^{-3}$, \citealt{1987MNRAS.224..801H}),  we obtain an estimate of
$E(B-V)\approx 0.4\pm$0.04. These  agree with the value yielded from SED fitting in \S2.2 \zhy{within errors}. The intensity ratios of narrow
hydrogen lines are also much steeper than that typically found in
the narrow line region (NLR) of AGNs \citep{2009ASPC..408..281Z} and star-forming
galaxies (\citealt{2012MNRAS.421..486X} and references therein), suggesting that
narrow emission lines might be heavily reddened by interstellar dust
clouds \zhy{ as well}. However, the relative measurement errors of narrow hydrogen lines are much larger than
that of broad lines due to the data quality available. High S/N and
resolution \zhy{ spectra are}  needed to confirm this.

\zhy{The emission-line  properties of SDSS J0802+5513 are typical for an NLS1, with narrow
Balmer emission line $FWHM({\rm H}\beta)\approx1800~\vel$,
strong optical Fe~II emission $R_{4570}\equiv
\frac{{\rm Fe~II\ }\lambda\lambda4434-4684}{{\rm H}\beta}\approx 1$, and
weak [O~III] emission $\frac{[{\rm O~III}]}{{\rm H}\beta}\approx
1.5$. The black hole
mass acquired is $M_{\rm BH}\approx 2.4\times10^8~M_{\odot}$ using the
empirical mass-luminosity-line width relation calibrated through
reverberation mapping \citep{2006NewAR..50..796P}.  In the most conservative case, we
integrate the infrared photometries from 2MASS K to WISE w4 to get a
lower limit of bolometric luminosity $L_{\rm
  bol}=1.2\times10^{46}~$erg~s$^{-1}$. Here we assume the radiation in
this wavelength range is from hot dust heated solely by the quasar
nucleus given the similar between infrared SED and the quasar
composite, and the dust has a full coverage of the central engine.
Here we did not include the emission beyond WISE w4.  A more
reasonable estimate is at least twice the value considering that FIR
is not included and covering factor of the dusty torus is about 0.5
based on fraction of obscured AGNs
\citep[e.g.][]{2005ApJ...620..629D,2008A&A...490..905H}. Alternatively, we can estimate the
bolometric luminosity using the monochromatic luminosity of  $\lambda
L_{\lambda5100}\approx 1.8\times10^{45}$~erg~s$^{-1}$ at 5100 \AA\
. Using the bolometric correction of $L_{\rm bol}/\lambda
L_{\lambda5100}=12$ \citep{2006ApJS..166..470R}, the corresponding
bolometric luminosity is $L_{\rm bol}\approx
2.2\times10^{46}~$erg~s$^{-1}$ and $L_{\rm bol}\approx
6.4\times10^{46}$~erg~s$^{-1}$, respectively, before and after
extinction correction. The inferred Eddington ratio is
$\dot{m}\sim0.4$--$1.4$, indicating the central SMBH is undergoing a rapid growth.  Assuming a mass-to-energy conversion
efficiency of $\eta =0.1$, we inferred a mass accretion rate of $\dot{M}_{\rm acc}\sim
2.1$--$7.7~M_{\odot}$~yr$^{-1}$}

\subsection{Absorption Lines Measurement \label{sec:absorption}}
\subsubsection{Normalization of Absorption Line Spectra}
\begin{figure}[!ht]
\includegraphics[scale=0.45]{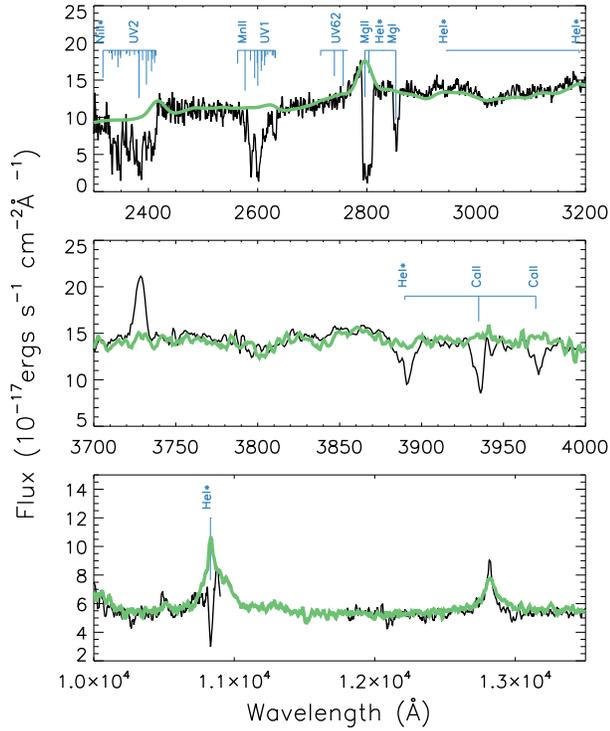} \caption{Expanded view of the observed
spectrum of SDSS J0802+5513 (black) in 3 absorption line regimes of
interest overplotted with model absorption-free spectrum (green).
Identified transitions are plotted as vertical bars on top of each
panel, with the bar lengths proportional to
log$(gf)$.\label{fig:abs_region}}
\end{figure}

To normalize the absorption line spectrum, absorption-free spectrum
of SDSS J0802+5513 must be recovered first. Two technique are
commonly adopted to achieve this goal, namely spectral decomposition
\citep[e.g.][]{2008ApJ...680..858L} and template match method \citep[e.g.][]{2006ApJ...639..716Z}.
 \zhy{We
split the observed spectrum into three regimes and use different approaches to reconstruct absorption-free spectra,
according to the different emission and absorption characteristics.}
The 3 spectrum regimes of interest are (1) Fe II+Mg II
regime (2100--3200 \AA ), (2) He I* $\lambda$3889+Ca II regime
(3700--4000 \AA ), and (3) He I* $\lambda$10830 regime (1--1.4
$\mu$m). The observed spectrum of the 3 regimes are bloated and
displayed in Figure \ref{fig:abs_region} with the best-fitted models
overlaid. Identified lines are labeled  \zhy{ by} vertical bars on the top
of each panel in the Figure and \zhy{listed in}  Table \ref{tbl:transition}. 

\zhy{Fe II+Mg II regime} We \zhy{employed} the spectral decomposition technique for this regime. The observed data were fitted by the combination of three components: a power law, a broadened Fe II+Fe III template and a broad Mg II emission line. The three components were reddened by the same SMC-like dust with $E(B-V)$ as a free parameter. The Fe II template built by \citet{2006ApJ...650...57T} is adopted, and the FeIII UV47 multiplet template is from \citet{2001ApJS..134....1V}. We assumed that Mg II broad line has the same profile and redshift as Balmer broad lines obtained in \S3.1. The best-fitted value of $E$(B-V)$\approx 0.30$ is consistent with that inferred from broad band SED fitting ($E$(B-V)$\approx 0.36$, \S2.2) and that estimated from broad hydrogen line ratios ($E$(B-V)$\approx 0.31$--$0.4$, \S3.1).

\zhy{He~I* $\lambda$3889+Ca II regime} We used the template matching method for this regime due to the fact that emission features in this regime, mainly arising from Fe~II, Ti~II and Cr~II etc., are very complex, and no appropriate templates are available for them \citep{2004A&A...417..515V,2006A&A...451..851V}. Furthermore the ratios of optical Fe II multiplets vary dramatically from object to object \citep{2001ApJS..134....1V}, Fe II templates built from in I Zw1 alone \zhy{cannot} fit this region well. We choose {\color{blue}} the observed spectra of Fe II strong quasars as templates to match the spectrum in this regime. The templates are chosen from DR 7 quasars with $EW_{\rm Fe II\ 4570}>60$ \AA\ and a median $S/N>20$ in  [O II] region. The best matched template  \zhy{ is the spectrum of SDSS J100446.52+600336.1 (as seen in Figure \ref{fig:abs_region}).}

\zhy{He~I* $\lambda$10830 regime} The He I* \zhy{$\lambda$10830}+Pa$\delta$ emission blends are seriously affected by strong He~I* $\lambda$10830 absorption line and the red wing of the blends falls at the gap between J and H bands. As  seen in Figure \ref{fig:abs_region}, the NIR composite \zhy{spectrum of quasar derived by} \citet{2006ApJ...640..579G} matches \zhy{ the observed} spectrum \zhy{ in this} regime quite well.
\begin{figure}[!ht]
\hspace*{0 cm}\includegraphics[scale=0.45]{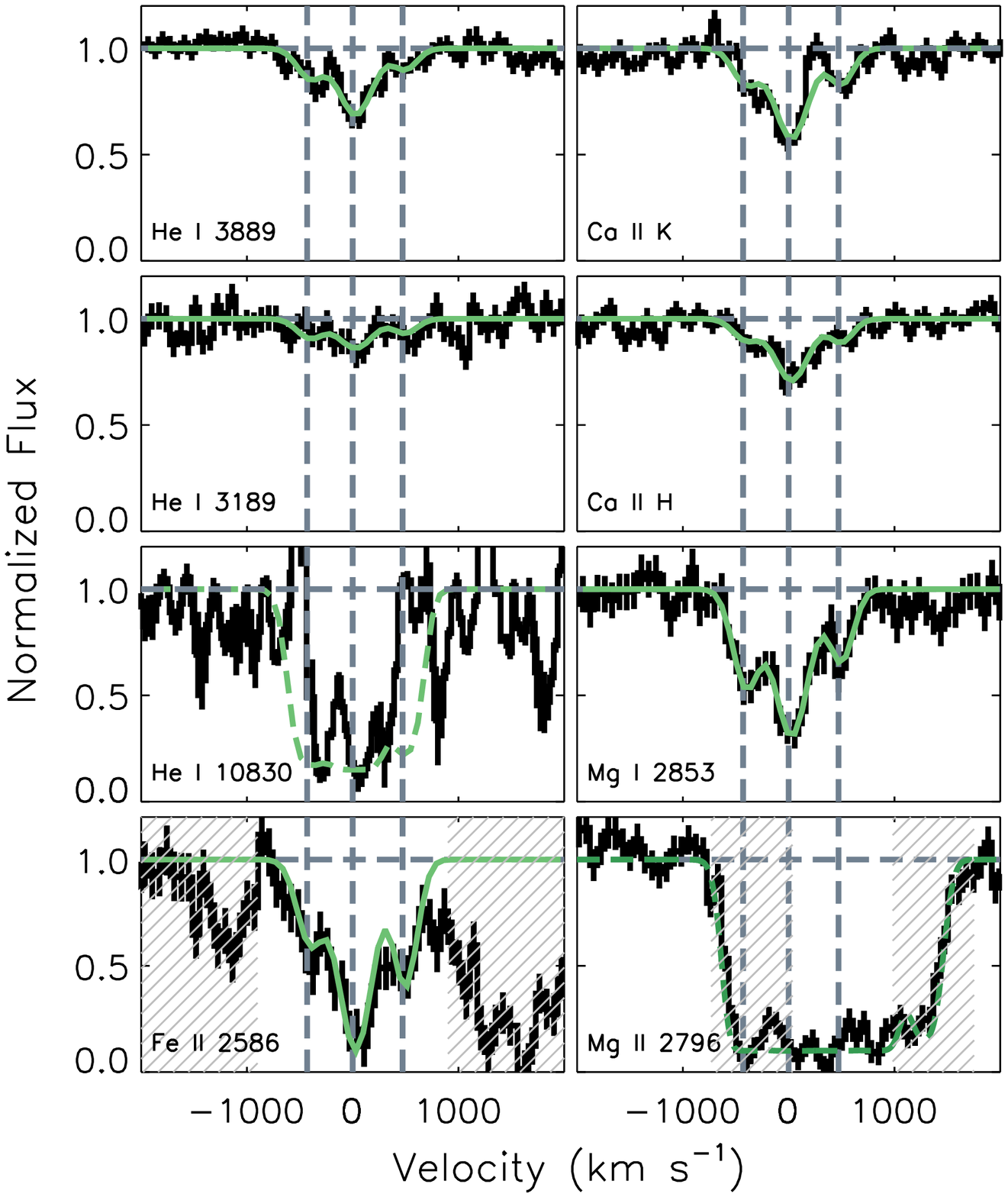}
\caption{Absorption spectrum of isolated lines plotted in a common
velocity space. The origin of the velocity scale is set to the
systematic redshift of the quasar. Observed data are shown in black curves with best-fitted models overlaid in green. The green dashed line in the right bottom
panel is the predicted profile of Mg II doublet using a column
density of $N_{\rm Mg\ II}=15.3$ cm$^{-2}$; and the green dashed line
in the right second panel from the bottom is the fitting of He I*
$\lambda$3889 scaled by 23.3, assuming a partial coverage factor of
0.85 (see \S\ref{sec:models} for detail). The Fe II* line blends
near Fe II $\lambda$2586 in the left bottom panel are shaded for
clarity. \zhy{Note that Mg II $\lambda$ is in fact a doublet with a velocity separation of 770 km s$^{-1}$. The zero velocity is  set for Mg II  $\lambda$2796, which is highly blended with Mg II $\lambda$2803. The left most 770 km s$^{-1}$ is solely contributed by Mg II $\lambda$2796, while the right most 770 km s$^{-1}$   by Mg II $\lambda$2803 (shaded for clarity). The region between the two shadowed area is contributed by both  lines.}\label{fig:isolated}}
\end{figure}

\begin{figure}[!ht]
\includegraphics[scale=0.45]{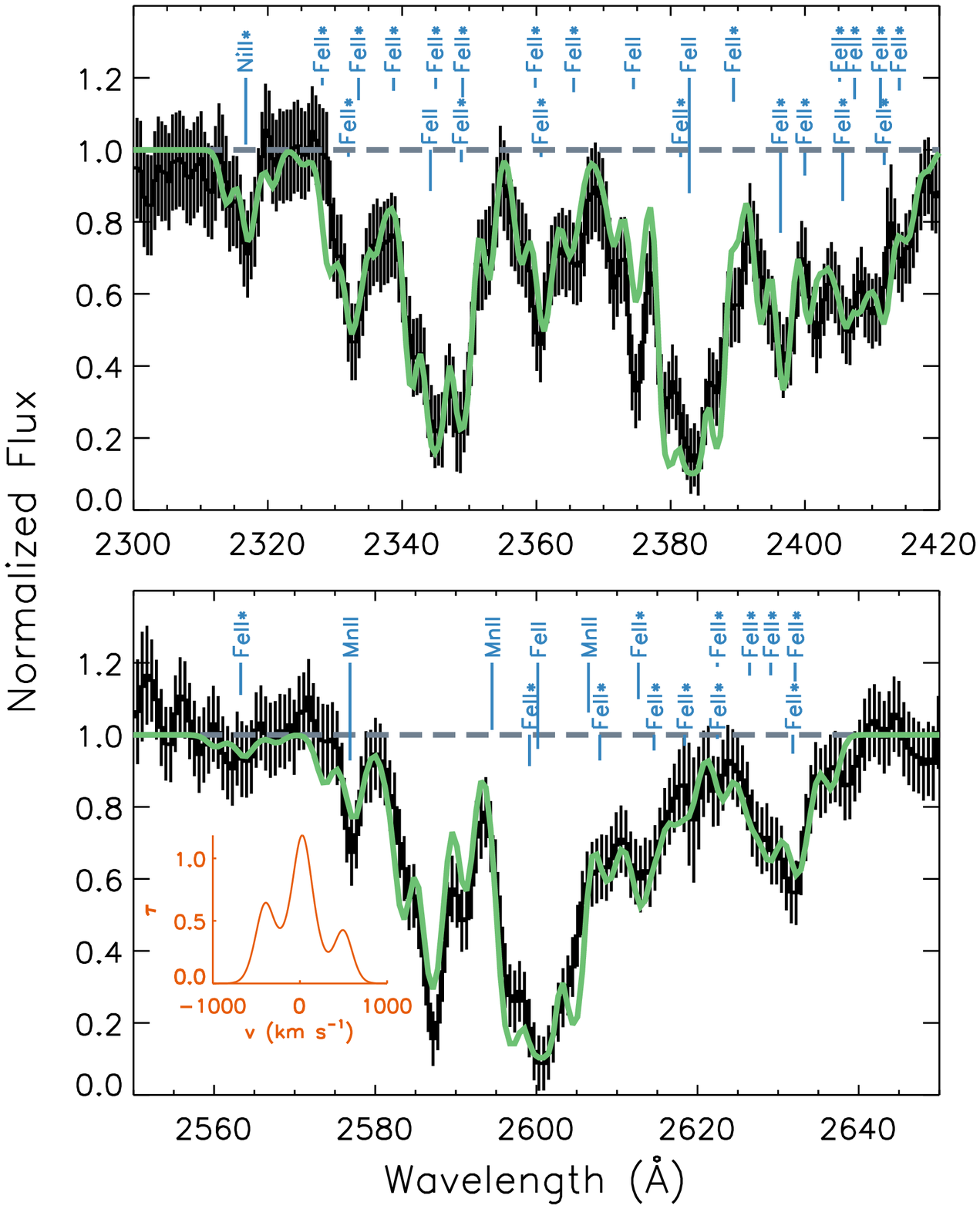}
\caption{Comparison between the observed data (black curves with
error bars) and the best-fit models (green) in the $\lambda$2350 and
$\lambda$2600 absorption blends. Identified transitions in Fe II, Mn
II and Ni II are labeled as in Figure \ref{fig:abs_region} (see also Table
\ref{tbl:transition} for the line list and related information). We
assume that all of the absorption lines have the same redshift and
profile as that of the mean of the isolated absorption lines (see Figure \ref{fig:isolated}), and the relative strengths are fixed to their theoretical
values. The assumed absorption line profile is shown as the
velocity-dependent curve of optical depth in the orange insert.
\label{fig:blended} }
\end{figure}

{The detected absorption lines (c.f. Table \ref{tbl:transition}) fall into two categories: relatively
isolated lines (Figure \ref{fig:isolated}) and heavily blended lines (Figure \ref{fig:blended}).} 
The normalized absorption line spectrum  is \zhy{ derived by dividing the observed spectrum by the 
absorption-free spectrum recovered above straightforwardly.}  This
normalization scheme is \zhy{ based} on the assumption that the absorption gas
 {covers both the continuum
source and BLR of quasar}. The validity of such an
assumption can be justified by \zhy{ checking the} residual flux at the
centroid of strong absorption lines.  The
residual fluxes at the centroids of Mg~II, Fe~II $\lambda$2600 and
He~I* $\lambda$10830 lines are so small that it would yield negative values
at these wavelengths if we subtracted the broad emission lines. This
implies that the absorber at least covers a significant part, if not
all of the BLR. \zhy{ In this case, the continuum source must be fully covered,}  since the size of the BLR is
about two orders of magnitude larger than that of the accretion disk.
{Indeed}, the apparent optical depth ratios of both Ca~II K, H and
He~I* $\lambda \lambda$3189, 3889 doublets support the assumption of
full coverage. The
column densities of Ca II and He I* evaluated by individual line of
the doublets based on the full coverage assumption \zhy{agree with each other within errors} (see Table \ref{tbl:isolated} and
detail in \S3.2). Note that the full coverage assumption is
consistent with the location of the absorption gas estimated in \S4,
which is $\sim 200$ pc from the central SMBH, about three orders of
magnitude larger than the BLR.

\subsubsection{Measurements of Absorption Lines}
We measured \zhy{the column densities of isolated lines and blended lines
 using different schemes, respectively}.
\begin{deluxetable}{lrccccc}
\tabletypesize{\scriptsize}
\tablecaption{Absorptions lines identified in SDSSJ0802+5513\label{tbl:transition} }
\tablewidth{0pt}
\tablehead{
\colhead{Wavelength(\AA)}&\colhead{log$(gf)$}&\colhead{Ion}
&\colhead{$E_{\rm low}$}&\colhead{$g_{\rm low}$}&\colhead{$E_{\rm up}$}&\colhead{$g_{\rm up}$}}
\startdata
10830.80 &-0.04  &\ion{He}{1}*  &159856 & 3& 169086 & 9\\
3889.80 &-0.72  &\ion{He}{1}*  &159856 & 3& 185565 & 9\\
3188.69 &-1.16  &\ion{He}{1}*  &159856 & 3& 191217 & 9\\
2852.97 & 0.270 &\ion{Mg}{1}   &     0 & 1&  35051 & 3\\
2796.36 & 0.100 &\ion{Mg}{2}   &     0 & 2&  35761 & 4\\
2803.54 &-0.210 &\ion{Mg}{2}   &     0 & 2&  35669 & 2\\
3934.83 & 0.134 &\ion{Ca}{2}   &     0  &2  &25414  &4\\
3969.65 &-0.166 &\ion{Ca}{2}   &     0  &2  &25192  &2\\
2576.87 & 0.433 &\ion{Mn}{2}   &     0  &7  &38807  &9\\
2594.49 & 0.270 &\ion{Mn}{2}   &     0  &7  &38543  &7\\
2606.46 & 0.140 &\ion{Mn}{2}   &     0  &7  &38366  &5\\
2344.2139&  0.057& \ion{Fe}{2}  &      0& 10&  42658&  8\\
2374.4612& -0.504& \ion{Fe}{2}  &      0& 10&  42115& 10\\
2382.7652&  0.505& \ion{Fe}{2}  &      0& 10&  41968& 12\\
2586.6500& -0.161& \ion{Fe}{2}  &      0& 10&  38660&  8\\
2600.1729&  0.378& \ion{Fe}{2}  &      0& 10&  38459& 10\\
2333.5156& -0.206& \ion{Fe}{2}*&    385&  8&  43239&  6\\
2365.5518& -0.402& \ion{Fe}{2}*&    385&  8&  42658&  8\\
2389.3582& -0.180& \ion{Fe}{2}*&    385&  8&  42237&  8\\
2396.3559&  0.362& \ion{Fe}{2}*&    385&  8&  42115& 10\\
2599.1465& -0.063& \ion{Fe}{2}*&    385&  8&  38859&  6\\
2612.6542&  0.004& \ion{Fe}{2}*&    385&  8&  38660&  8\\
2626.4511& -0.452& \ion{Fe}{2}*&    385&  8&  38459& 10\\
2328.1112& -0.684& \ion{Fe}{2}*&    668&  6&  43621&  4\\
2349.0223& -0.269& \ion{Fe}{2}*&    668&  6&  43239&  6\\
2381.4887& -0.693& \ion{Fe}{2}*&    668&  6&  42658&  8\\
2399.9728& -0.148& \ion{Fe}{2}*&    668&  6&  42335&  6\\
2405.6186&  0.152& \ion{Fe}{2}*&    668&  6&  42237&  8\\
2607.8664& -0.150& \ion{Fe}{2}*&    668&  6&  39013&  4\\
2618.3991& -0.519& \ion{Fe}{2}*&    668&  6&  38859&  6\\
2632.1081& -0.287& \ion{Fe}{2}*&    668&  6&  38660&  8\\
2338.7248& -0.445& \ion{Fe}{2}*&    863&  4&  43621&  4\\
2359.8278& -0.566& \ion{Fe}{2}*&    863&  4&  43239&  6\\
2405.1638& -0.983& \ion{Fe}{2}*&    863&  4&  42440&  2\\
2407.3942& -0.228& \ion{Fe}{2}*&    863&  4&  42401&  4\\
2411.2433& -0.076& \ion{Fe}{2}*&    863&  4&  42335&  6\\
2614.6051& -0.365& \ion{Fe}{2}*&    863&  4&  39109&  2\\
2631.8321& -0.281& \ion{Fe}{2}*&    863&  4&  38859&  6\\
2345.0011& -0.514& \ion{Fe}{2}*&    977&  2&  43621&  4\\
2411.8023& -0.377& \ion{Fe}{2}*&    977&  2&  42440&  2\\
2414.0450& -0.455& \ion{Fe}{2}*&    977&  2&  42401&  4\\
2622.4518& -0.951& \ion{Fe}{2}*&    977&  2&  39109&  2\\
2629.0777& -0.461& \ion{Fe}{2}*&    977&  2&  39013&  4\\
2332.00  & -0.720& \ion{Fe}{2}*&   1873& 10&  44754&  8\\
2348.81  & -0.470& \ion{Fe}{2}*&   1873& 10&  44447&  8\\
2360.70  & -0.700& \ion{Fe}{2}*&   1873& 10&  44233& 10\\
2563.30  & -0.050& \ion{Fe}{2}*&   7955&  8&  46967&  6\\
2715.22  & -0.440& \ion{Fe}{2}*&   7955&  8&  44785&  6\\
2740.36  &  0.240& \ion{Fe}{2}*&   7955&  8&  44447&  8\\
2756.56  &  0.380& \ion{Fe}{2}*&   7955&  8&  44233& 10\\
2316.72  &0.268 &\ion{Ni}{2}*  & 8394 &10  &51558 & 8\\

\enddata
\end{deluxetable}
\zhy{ The isolated} He~I* $\lambda\lambda$3189, 3889, Fe~II $\lambda$2586,
Ca~II K, H, and Mg~I lines show nearly identical velocity structure \zhy{ as seen in Figure \ref{fig:isolated}}.
Each line has three distinct components, as indicated by the vertical
dashed lines. The deepest component is centered at the
quasar systematic redshift. Two shallower components are
symmetrically distributed around the deepest one with a velocity
shift of $\Delta v\sim \pm$500 km s$^{-1}$. Assuming the background
source fully covered and the absorption lines \zhy{ moderately} resolved,
we evaluated the optical depth profiles of the 6 absorption lines as
$\tau (v)=$ -ln$I_{\rm r}(v) $, where $I_{\rm r}$ is the residual intensity of the
normalized spectrum. Each of the 6
\zhy{ lines was} fitted with 3 Gaussians. The width and centroid of the
corresponding Gaussian are tied during the fit. The best-fitted
optical depth as a function of velocity is shown in the insert in
Figure \ref{fig:blended}. The equivalent widths ($EW$s) of these 6 lines are measured
from the best-fitted models, which are overplotted in Figure
\ref{fig:isolated}. We also calculated the column densities of the
corresponding lines by integrating their best-fitted apparent
optical depth profiles. The measured $EW$s and column densities are
listed in Table \ref{tbl:cols}. Both of Mg II doublet and He I* $\lambda$10830 lines are seriously saturated and no direct
measurement is available. Their absorption troughs are flat-bottomed
with residual flux of $\lesssim $ 10\% at the deepest points
\footnote{Either partially covered background light, scattered light
of AGN, or starlight from the host galaxy could contribute to the
residual flux. Current available data quality are not of enough to
distinguish these possibilities. He I* $\lambda$10830 is seriously
affected by sky line residuals.}. As a conservative estimate,
assuming a covering factor of 85\%, we \zhy{ rescaled} the best-fitted He
I* $\lambda$3889 optical depth profile by a factor of 23.3
($f\lambda$ ratio of $\lambda$10830 to $\lambda$3889, \citealt{2011ApJ...728...94L}) to
generate a He I* $\lambda$10830 absorption line model. The model is
overlaid on the observed data in Figure \ref{fig:isolated}. Mg II $\lambda$2796 is
seriously blended with Mg II $\lambda$2803, we used the best-guessed
Mg$^+$ column density from photoionization simulation in \S4 to
create the model over-plotted in Figure \ref{fig:isolated}. In Table \ref{tbl:isolated}, we take half
of the integrated $EW$ of the blend as a rough estimate for each of
the doublet.

\begin{deluxetable}{lrrcc}
\tablecaption{Ionic column densities for isolated lines in SDSS J0802+5513.\label{tbl:isolated}}
\tablewidth{0pt}
\tablehead{
\colhead{Species} & \colhead{transtions}&\colhead{$EW$ (\AA)}&\colhead{$N_{\rm ion}$ (cm$^{-2}$)}
}
\startdata
Ca II&3969.59&2.15$\pm$0.16&13.59$\pm$0.07\\
Ca II&3934.78&3.52$\pm$0.17&13.51$\pm$0.05\\
He I*&3889.74&2.54$\pm$0.18&14.73$\pm$0.07\\
He I*&3188.66&0.82$\pm$0.14&14.82$\pm$0.17\\
Mg I&2852.96&4.51$\pm$0.53&13.53$\pm$0.11\\
Mg II&2803.53&5.82$\pm$0.21\tablenotemark{a}&\nodata\\
He I*&10830.40&$>$29.63&\nodata
\enddata
\tablenotetext{a}{Half of the total equivalent width of Mg II
doublet absorption lines.}
\end{deluxetable}

\zhy{HiBALs like C IV, Si IV are often much stronger and wider than
  LoBALs like Mg II, Al III \citep{2010ApJ...714..367Z,2014ApJ...791...88F}}, it is remarkable that He~I*, Fe~II, Ca~II and Mg~I lines, \zhy{which are arising from ions with very different ionization potentials,} have
almost the same velocity structure, \zhy{and this indicates that SDSS
J0802+5513 may have an origin other than traditional BALs}. He~I* $\lambda\lambda$3189, 3889, 10830 arise from metastable triplet
level He I 2$^3$S at a rather high excitation energy of 19.6 eV. The
level is populated by recombination from He$^+$ ions \citep{2012RAA....12..369J}, which are created by photons with energies of $h\nu > 24.56$
eV and are destroyed by photons with $h\nu > 54.42$ eV. They survive
in much different conditions than that of Ca$^+$ ions and neutral Mg
atoms that give rise to Ca II K, H and Mg~I lines. Mg atoms are
destroyed by photons with $h\nu > 7.65$ eV; and Ca$^+$ ions are
created by photons with energies of $h\nu > 6.11$ eV and are
destroyed by photons with $h\nu
> 11.87$ eV. The nearly identical profile of He~I*,
Ca~II and Mg~I lines implies that all of the rest detected
absorption lines should have the same velocity structure, because all \zhy{ of} them originate from singly ionized
ions with surviving conditions in between that of He$^+$ and of
Ca$^+$ and neutral Mg.

\begin{deluxetable}{lrcccc}
\tablecaption{Ionic column densities for the blended lines in
SDSS0802+5513.\label{tbl:cols}} \tablewidth{0pt} \tablehead{
\colhead{Species} & \colhead{E (cm$^{-1}$)}&
\multicolumn{2}{c}{{log$_{10}$N (cm$^{-2}$)}}}\startdata
\ion{Fe}{2} &     0& 14.76$\pm$0.10  \\
\ion{Fe}{2*} & 385  & 14.11$\pm$0.17 \\
\ion{Fe}{2*} & 668  & 13.92$\pm$0.29 \\
\ion{Fe}{2*} & 863  & 13.79$\pm$0.40  \\
\ion{Fe}{2*} & 977  & 13.52$\pm$0.56  \\
\ion{Fe}{2*} &1873   & 15.11$\pm$0.30\tablenotemark{a}  \\
\ion{Fe}{2*} &7955   & 13.27$\pm$1.89  \\
\ion{Fe}{2} &Total  &15.34$\pm$0.5 \\
\ion{Ni}{2*}  & 8394  &  13.85$\pm$0.52\\
\ion{Mn}{2} &0  & 13.31 $\pm$0.47\\
\enddata
\tablenotetext{a}{The column of this level seems strangely
large compared to that of ground level, but similar result is seen in
QSO 2359-1241, see \citet{2008ApJ...688..108K} }
\end{deluxetable}

For the Fe II UV1 and UV2 regimes displayed in Figure
\ref{fig:blended}, the absorption lines are too heavily blended to fit them
separately. As listed in Figure \ref{tbl:transition}, We identified Fe II lines that arise from ground levels
and from excited levels up to 7955 cm$^{-1}$. Ni II$^*$ lines from the excited level of
wavelength number 8395 cm$^{-1}$ and Mn II lines from the ground
level are also identified. All of the identified absorption lines
were fitted simultaneously using the same  \zhy{ optical depth profile} generated above
from the isolated lines \zhy{(shown in the insert of Figure \ref{fig:blended})}.
We also calculated $EW$s of  \zhy{ individual} absorption lines by
integrating \zhy{the normalized flux of} their models. The best-fitted column densities are
summarized in Table \ref{tbl:cols} and the best-fitted model is compared with the
observed absorption line spectrum in Figure \ref{fig:blended}.
\begin{figure}[!ht]
\centering
\includegraphics[scale=0.45]{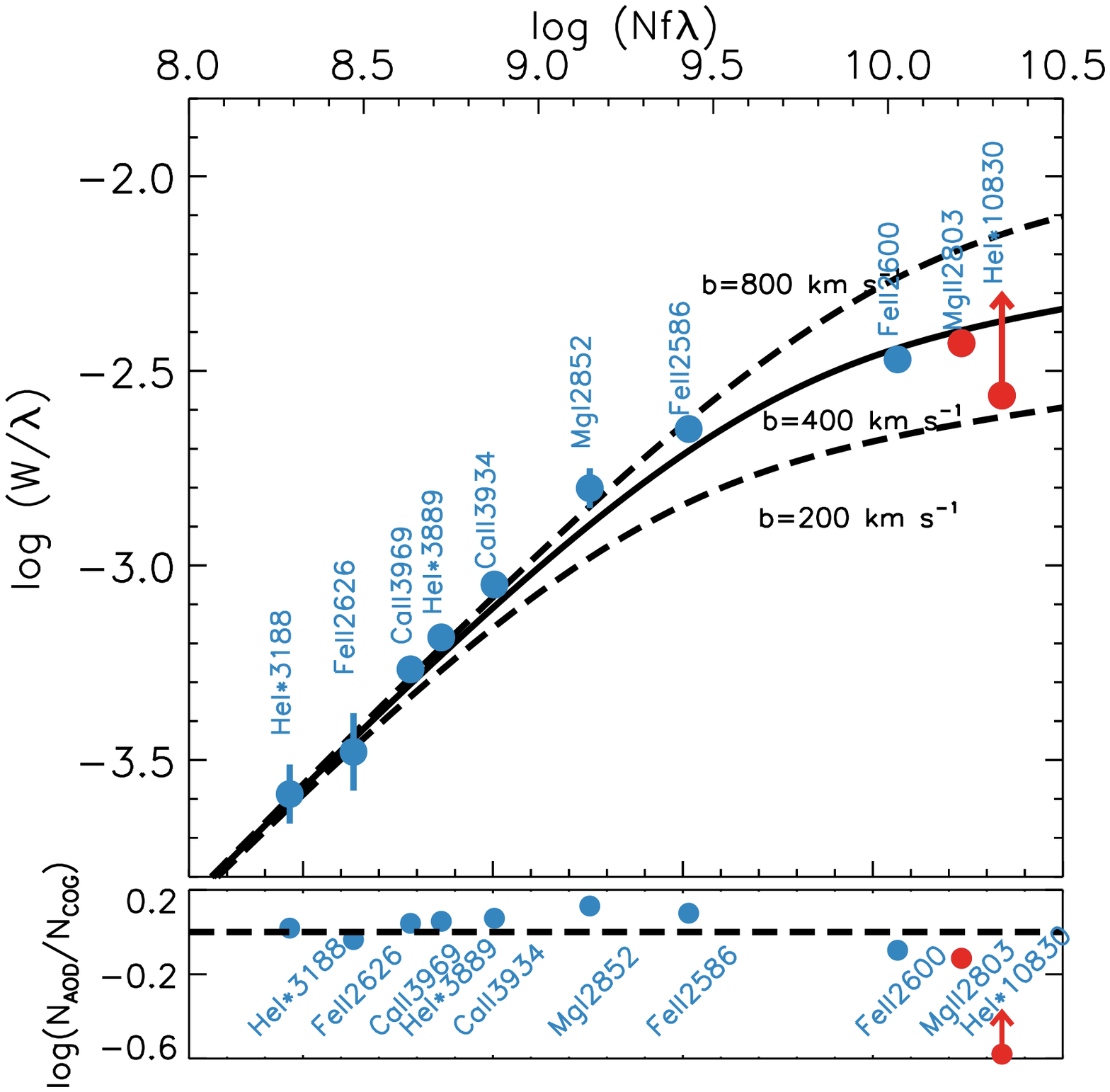}
\caption{Comparison between apparent optical depth (AOD) and curve
of growth (COG) analysis. In the upper panel, AOD measurements of
isolated lines are shown in blue filled circles (the inferred values
of He I* $\lambda$10830 and Mg II $\lambda$2803 are shown in red.
See Figure \ref{fig:isolated} and \ref{fig:blended} for the line measurements.); and the best-fit COG
using ground level Fe II lines is shown in filled black curve ($b=400
\vel$). Also plotted there in dashed black curves are COGs
with $b=200$ and $800\vel$ to guide the eyes. It can be seen in
the lower panel that the differences between AOD and COG column
density measurements for most ions are less than 0.1 dex.\label{fig:cog} }
\end{figure}
This  model recovers the observed data very well. This implies
that the apparent optical depth (AOD) method, which was adopted to
 \zhy{ measure the column densities}, is reasonable.

As pointed out by \citet{1986ApJ...304..739J}, large populations of
absorption lines can be analyzed collectively using the standard,
single-component  \zhy{ curve of growth (COG) method}.
We combine the 3 components in SDSS J0802+5513 and perform such
single-component COG analysis as a double-check to the AOD
measurements. Five Fe~II lines with $gf$ span of about 1 dex (Table
\ref{tbl:transition}) were used to evaluate the best-fitted COG. We calculated
theoretical COGs of various $b$ values and searched for the best
match  the measured $EW$s of the ground
level Fe II lines. Both of $b$ and $N_{\rm Fe\ II}$ are free parameters during the fit.
The best-fitted COG of Fe~II lines is presented in Figure \ref{fig:cog}.
We interpolated the $EW$ measurements on the best fitted COG to infer
COG column density {for other transitions}.
The  {differences} of column densities evaluated by the AOD and COG methods,
log$(N_{\rm AOD}/N_{\rm COG})$, are plotted in the lower panel of Figure \ref{fig:cog}.
The {differences} are  {negligible within errors,
for most lines, {in order of 0.1 dex.} The AOD method relies on two assumptions: (1) the absorption lines
are completely resolved, and (2) the absorber fully covers the
background emission source. The overall agreement between  \zhy{$N_{\rm AOD}$ and $N_{\rm COG}$}
indicates that the two assumptions are at least an acceptable approximation.

\section{Physical Conditions and Location of the Absorption Gas}
In this section, we  explore the physical conditions of the
absorption gas and locate the gas with the aid of photoionization
model calculations and using the column densities of various
ions/levels reliably measured in \S3.2. Fe II* absorption lines that
arises from excited levels \zhy{ are sensitive} to  the
electron density $n_{\rm e}$ (e.g. \citealt{2001ApJ...546..140A}; \citealt{2010ApJ...709..611D}) \zhy{ in absorbers}. \zhy{ Moreover,} He I* lines are a good diagnostics for
constraining ionization parameter $U$, which is defined as
\begin{equation}
\label{equ:U} U = \frac{1}{4\pi r^2 n_{\rm H}c} Q = \frac{1}{4\pi
r^2 c n_{\rm H}} \int_{\nu_0}^{\infty}\frac{L_\nu }{h\nu} d\nu,
\end{equation}
where $\nu_0$ is the frequency corresponding the hydrogen edge, and
$Q$ is the emission rate of hydrogen ionization photons. Once $U$ and
$n_{\rm e}$ are well constrained, the distance of absorption gas can
be inferred from Equation (\ref{equ:U}). We carry out
detailed analysis using a photoionization model in \S4.1, and  discuss
in \S4.2 the possible dependence on metal abundances, SED of the
background quasar, and dust reddening effects. We  \zhy{ run} photoionization code CLOUDY to
carry out the model simulations (version c13.00; c.f. \citealt{1998PASP..110..761F}).

\subsection{Basic Model\label{sec:models}}
First we will show that the column density ratio $N_{\rm Fe\
II}/N_{\rm He\ I*}$ observed in SDSS J0802+5513 requires the
absorber to be thick enough to have a partially-ionized or neutral
zone behind the hydrogen ionization front.
We started by considering a gas slab with a density of $n_{\rm H}$ and a total column density of
$N_{\rm H}$, which is illuminated by quasar radiation (SED from \citealt{1987ApJ...323..456M},
hereafter MF87). We calculated a grid of models  \zhy{by varying} $U$, $n_{\rm H}$, and $N_{\rm H}$, 
\zhy{with metallicity fixed to solar abundance}. As an
example, we plot one \zhy{result} of the model calculations in Figure \ref{fig:ion_structure}. The
ionization structure is shown as a function of the depth parameter
$N_{\rm H}$ from the illumination surface.  \zhy{ Transitions from} different ionization
state behaves differently in the $N_{\rm ion}-N_{\rm H}$ plane, \zhy{ yielding}
a sensitive dependence of ion column density ratios on $N_{\rm H}$. This
implies that these metal ion column density ratios are good
constraints to the absorber thickness. Specifically, only the very
narrow $N_{\rm H}$ range, labeled {by two vertical dashed lines} in Figure \ref{fig:ion_structure},
can  \zhy{produce} the value of
log$(N_{\rm Fe\ II}/N_{\rm He\ I*})=0.60\pm0.35$ observed in SDSS
J0802$+$5513 for the adopted $U$ and $n_H$. 
This is understandable considering that Fe~II (and other singly-ionized metal
ions or neutral atoms) and He I* 2$^3$S survive at different conditions.
\begin{figure}[!ht]

\includegraphics[angle=-90,scale=0.34]{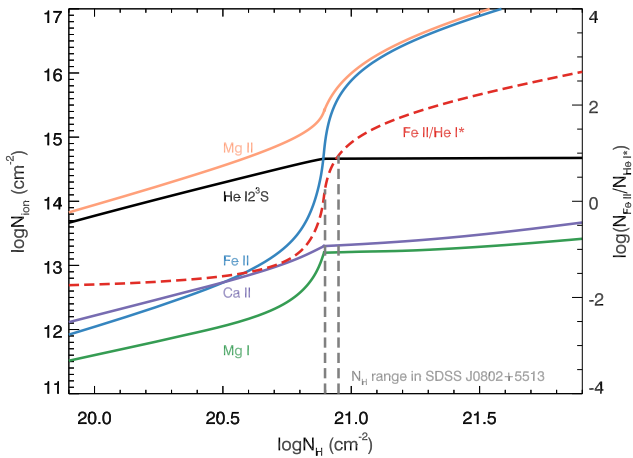}
\caption{Ionic column densities as a function of the depth parameter
$N_{\rm H}$ for a specific photoionization model with $U=10^{-2}$
and $n_{\rm H}=10^5$ cm$^{-3}$. The incident quasar SED is from
Mathews \& Ferland (1987, MF87) and a solar abundance is assumed.
Note the different behavior of different ions. The sensitive
dependence of the column density ratio $N_{\rm Fe\ II}/N_{\rm He\
I*}$ on $N_{\rm H}$ (the right-y axis with a different set of tick
marks) constrains the total column density of SDSS J0802+5513 to a
rather narrow range as labeled by the two vertical dashed lines for
the specific model (see \S4.1 for the model detail).
\label{fig:ion_structure} }
\end{figure}
\zhy{He I* 2$^3$S state is mainly
populated by the recombination of a He$^+$ ion with an electron.} As $N_{\rm H}$ increases, the main
ionization state of helium changes from He$^{++}$ to He$^{+}$,
resulting in a sharp increase in He I* 2$^3$S near the ionization
front. {Further into the cloud}, the ionizing photons quickly run out
and helium becomes almost neutral, which in turn prevents the
generation of  He I* 2$^3$S ions. \zhy{However},
the existence of such an ionization front does not prevent the
formation of ions with lower ionization potentials, such as Ca II,
Mg II and Fe II. As a result, $N_{\rm Fe\ II}/N_{\rm He\ I*}$
continues to increase after the front. Therefore, once the gas
abundance and the incident SED are appointed, $N_{\rm Fe\ II}/N_{\rm
He\ I*}$ ratio can give a tight constrain to the thickness of the
absorber, provided the absorber is thick enough to generate the
maximum $N_{\rm He\ I*}$. More importantly, this $N_{\rm He\ I*}$
corresponds to a nearly unique value of $U$ for a large range of gas
density $n_{\rm H}$.
\begin{figure}[!htp]
\includegraphics[angle=-90,scale=0.35]{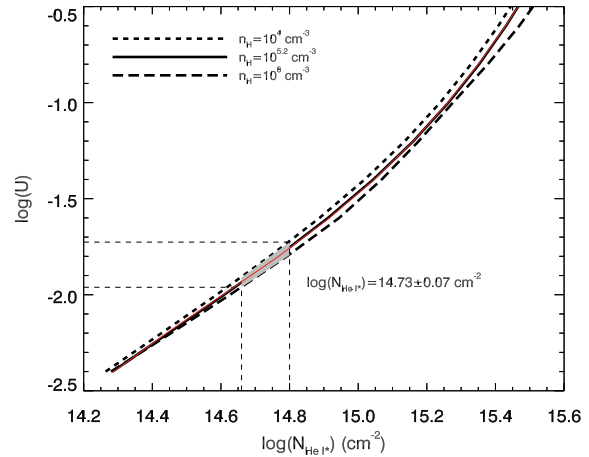}
\caption{Calculated He I* column densities of the photoionization
models with $n_{\rm H}\sim 10^{4}-10^{8}~$cm$^{-3}$ and $U\sim
10^{-2.5}-10^{0}$ (see \S4.1 for detail). Note that $N_{\rm H}$ is
strongly dependent on $U$ and insensitive to $n_{\rm H}$. The dashed lines
indicate the constraint $U$ range of SDSS J0802+5513.\label{fig:he_U}}
\end{figure}
As a demonstration, we have run an extensive grid of simulations
with log$U$ varying from -2.5 to 0 with a step of 0.1 and log$n_{\rm
H}$ varying from 4 to 8 with step of 0.2. \zhy{The upper limit of
$n_{\rm H}$, $10^{8}~$cm$^{-3}$, is determined by the fact that Balmer absorption lines,
arising from excited hydrogen $n$=2 level generated by collision, are
not detectable in SDSS J0802+5513 \citep{2011ApJ...728...94L}}.
The lower limit to the density can be set by equilibrium
equation of the lowest Fe II excited level 385 cm$^{-1}$ (de Kool et
al. 2001).
Using a two-level approximation,
\begin{equation}
\label{eq2} {n_1n_{\rm e}q_{12}=n_2n_{\rm e}q_{21}+n_2A_{21}}
\end{equation}
where the subscripts 1 and 2 represent ground level and the 375
cm$^{-1}$ level, respectively, and neglecting collisional term, we
\zhy{derive} a lower limit of $n_{\rm H}\gtrsim 10^4~$cm$^{-3}$. The stop
column density of $N_{\rm H}=10^{24}~$cm$^{-2}$ is so chosen to enclose a
fully developed ionization front for the largest $U$
concerned\footnote{H II region scales approximately as
log$_{10}N_{\rm H}\approx23+$log$U$}. In Figure \ref{fig:he_U}, we
show  the model prediction of $N_{\rm He\ I*}$ for a given set of
$U$ and $n_{\rm H}$. The photoionization parameter is constrained
to a very narrow range of $-2.0\lesssim {\rm log}U \lesssim -1.8$ by the
observed $N_{\rm He\ I}*$ \footnote{The model simulations show an
interesting fact that $N_{\rm He\ I}*$ is strongly dependent on $U$,
but it is insensitive to $n_{\rm H}$. This suggests that, for the
range of model parameters investigated, $N_{\rm He\ I*}$ could be a
good indicator to $U$. To a rough approximation ($\sim$ 0.1 dex
error), we may neglect the effect of $n_{\rm H}$ and adopt the
following relation to estimate $U$ using only $N_{\rm He\ I}*$,
\begin{equation}
\label{equ:u_n} {\rm log}U=0.50-(92.25-5.88~{\rm log}N_{\rm He\ I*})^{0.5}.
\end{equation}
This empirical relation is consistent with the model results of
well-studied individuals in the literature with He~I* and Fe~II*
measurements. E.g., for the intrinsic absorber in QSO 2359$-$1241,
detailed model calculations indicate a range of $-2.8\lesssim {\rm log}U
\lesssim -2.7$ \citep{2001ApJ...546..140A}, while the result from Equation
(\ref{equ:u_n}) is ${\rm log}U\approx-2.83$. See \S\ref{sec:systematics}
for detailed discussion on systematics including incident SEDs,
metallicities, and dust reddening effect.}.
\begin{figure}[!htp]
\vspace{0.5cm}\includegraphics[angle=-90,scale=0.35]{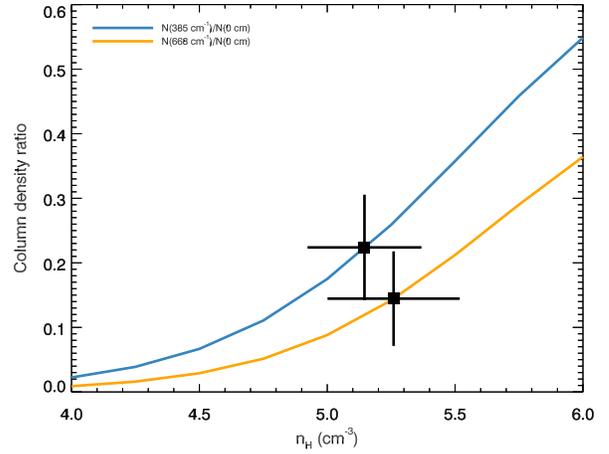}
\caption{Calculated dependence of Fe~II excited level populations
(385 cm$^{-1}$ and 668 cm$^{-1}$ relative to the ground level as
labeled by the legends) on gas density. The measurements are shown
as filled squares with 1-$\sigma$ statistical errors. \label{fig:density}}
\end{figure}
Using the median value of log$U=-1.9$, we calculated the column
density ratios of the excited to ground level of Fe$^+$ ions as a
function of $n_{\rm H}$. Figure \ref{fig:density}  \zhy{presents the
results of the 385 cm$^{-1}$ and 668 cm$^{-1}$ levels. The estimated density is  $n_{\rm H}\approx$10$^{5.2\pm0.3}$ cm$^{-3}$.} Measurement
 \zhy{uncertainties} of higher levels are too large to \zhy{estimate}
 \zhy{ the electron density reliably.}
 \zhy{Note that this density is  well within the 
aforementioned value range of $n_{\rm H}\sim10^{4}$--$10^8$ cm$^{-3}$.} We recalculated the model using the best
estimates of log$U=-1.9$ and $n_{\rm H}\approx$10$^{5.2}$ cm$^{-3}$.
The model calculation was stopped at  $N_{\rm H}$=10$^{21}$
cm$^{-2}$  where  the observed $N_{\rm Fe\ II}$ is reached. The model
column densities are compared with the observed values in Figure
\ref{fig:col_predict}. The observed column densities can be well
reproduced by the model for all ions but Ca~II, which is slightly
under-estimated. We normalized the MF87 SED to the observed
luminosity of SDSS J0802$+$5513 at the WISE w4-band ($\lambda_{\rm eff}
= 22~\mu m$), and obtained an estimate of the ionizing photons rate
of $Q_{\rm MF87} = 2.3 \times 10^{56}$ s$^{-1}$, which should not be
affected by reddening. Substituting $Q_{\rm MF87}$ and the best
evaluates of $U$ and $n_{\rm H}$ into Equation (\ref{equ:U}), we
inferred an estimate to the distance of the absorption gas from the
SMBH, $R\approx 200~(\frac{U}{10^{-1.9}})^{0.5}
(\frac{n_{\rm e}}{10^{5.2}{\rm \ cm}^{-3}})^{0.5}$ pc. The inferred physical
thickness of the absorber is $\Delta R \sim N_{\rm H}/n_{\rm H}
\approx 0.02$ pc, which is rather small compared to its distance
from the central engine, $\Delta R/R \sim 10^{-4}$.
\subsection{Effect of SED, Metallicity, and Dust\label{sec:systematics}}
Only the MF87 SED and a solar abundance were considered in the basic
model described above. Adopting different metallicities or incident
SEDs might introduce systematics to model calculations. Neither was
dust reddening taken into account in the calculations, and yet it is
in fact observed in SDSS J0802+5513. In order to assess these
possible systematics, we repeated the calculations adopting
different metallicities and incident SEDs, and incorporating the effect of dust (including its effect
 on heating and cooling in the photoionized gas, as calculated with Cloudy). Two dust configurations were considered here. (1) Dust is uniformly mixed with the absorption gas. We use the pre-stored
abundance set ``H II region with grains" offered by CLOUDY as
``Orion nebula dust", which is essentially the average condition of
the Orion Nebula (\citealt{1991ApJ...374..580B}; see the note of Table
\ref{tbl:models}). In this case, metals are depleted into dust and
the gas phase abundance is sub-solar. (2) The dust is in front of
the absorber (a dust screen case). In this case, the dust resides in
a very high ionized foreground gas \citep{2010ApJ...709..611D}, which does
not leave imprints on the observed spectra of the quasar but alters
the shape the incident continuum. In Figure \ref{fig:seds}, we show the MF87 SED
redden with $E(B-V)$=0.36 using the model extinction curve of ``SMC
bar" from \citet{2001ApJ...548..296W} since there is no observed
extinction curve available in the extreme ultraviolet (EUV). Also
displayed in the figure is the standard AGN SED in hazy document
(CLOUDY command: AGN T=1.5e5 K, a(ox)=-1.4, a(uv)=-0.5, a(x)=01).
This SED has a big blue bump peaked at a lower energy than MF87, and
we refer to it as ``SOFT SED".\footnote{We have also considered cases where ionizing sources are in situ with the gas, e.g. stellar sources. For solar abundance, SEDs from young stellar populations (e.g. 1 Myr SSP from \citealt{1999ApJS..123....3L}) would  result in similar results as SOFT SEDs do. In these cases, we \zhy{cannot} use this the infer the distance of gas from the central black hole, so we do not list these results in Table \ref{tbl:models} for consistency. Yet the fact itself, that the starburst SEDs could generate observed ionic column density pattern, would fit in our preferred picture of starburst-driven gas flow in SDSS J0802+5513 (see \S\ref{sec:origin} for detail)}

\begin{figure}[!htp]
\includegraphics[angle=-90,scale=0.4]{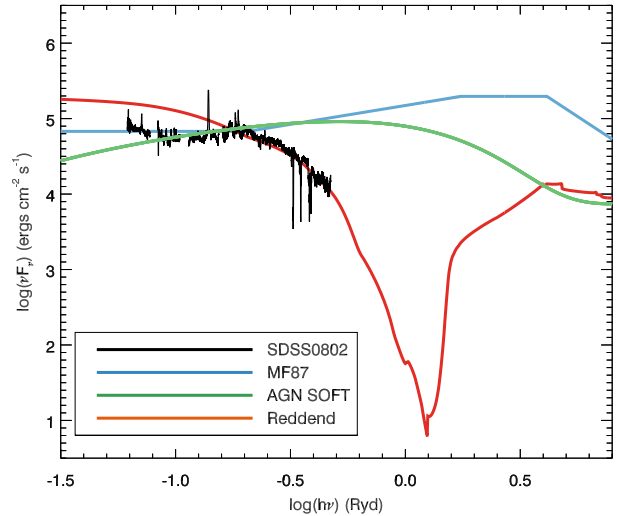}
\caption{Three incident SEDs adopted in the photoionization
simulations. The data are normalized to the spectrum of SDSS
J0802+5513 at $\lambda =5100$ \AA~in  quasar rest-frame (see \S4
for detail). \label{fig:seds}}
\end{figure}
In all cases, $U$ and $N_{\rm H}$ are adjusted to best match the
observed $N_{\rm Fe\ II}$ and $N_{\rm He\ I*}$, and $n_{\rm H}$ is
fixed to be 10$^{5.2}$ cm$^{-3}$. The results are summarized in
Table \ref{tbl:models} and are compared with observations in Figure
\ref{fig:col_predict}. It can be seen there that models assuming a
solar gas-phase abundance reproduce the observed column densities
quite well, while dusty-gas models assuming an intrinsic solar
abundance under-estimate $N_{\rm Ca\ II}$ and $N_{\rm Mg\ I}$. This
is expectable, since in these models the gas phase calcium and
magnesium are heavily depleted into dust grains. The best-fit models
yield Ca/H$=-7.7$ and Mg/H$=-5.5$, more than one order of magnitude
lower than the solar values. An intrinsic super-solar abundance is
needed for the dusty-gas models to compensate metal depletion. To
summarize, all of the acceptable models require $U\sim
10^{-2}-10^{-1.5}$ and $N_{\rm H}\sim 10^{21}-10^{21.5}$ cm$^{-2}$. The
distances inferred from all models circulated in Table
\ref{tbl:models} are in the range of $R\sim 100-250$~pc. Dusty-gas
models in general require larger $U$ and accordingly infer a smaller
distance compared with dust-free models.\begin{figure}[!htp]
\hspace*{-1cm}\includegraphics[angle=-90,scale=0.4]{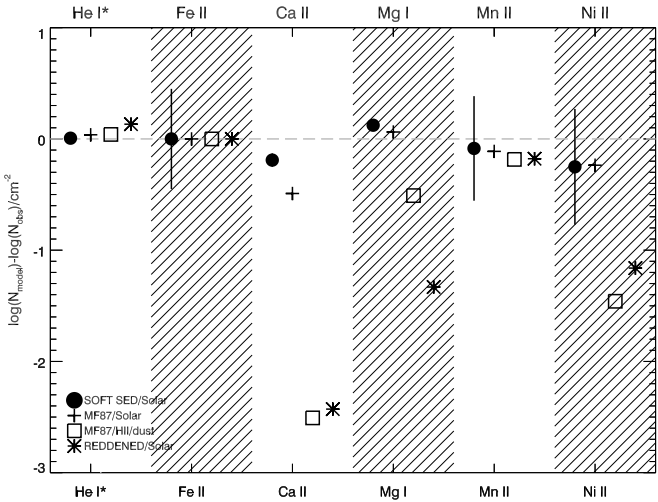}
\caption{Comparison between column density measurements and model
calculations (see \S4 and Table \ref{tbl:models} for the detail).
Measurement errors are only shown for the case of SOFT SED for
clarity. \label{fig:col_predict}}
\end{figure}
\begin{deluxetable}{lrrrrcrrrr}
\tablewidth{0pt}
\tablecaption{CLOUDY models\label{tbl:models}}
\tablehead{
\colhead{SED\tablenotemark{a}}           & \colhead{Z}      &
\colhead{U}          & \colhead{N$_{\rm H}$}  &
\colhead{He I*}          & \colhead{Fe II*}    &
\colhead{Ca II}  & \colhead{Mg I}  &
\colhead{Mn II} & \colhead{Ni II\tablenotemark{d}}
}
\startdata
observation&\nodata&\nodata& \nodata&   14.73&   15.34&   13.59&   13.53&   13.40&   15.34\\
MF87&HII/DUST\tablenotemark{b}&   -1.50&   21.49&   14.75&   15.34&   11.05&   13.00&   13.22&   13.89\\
MF87&solar\tablenotemark{c}&   -1.80&   21.21&   14.75&   15.34&   13.07&   13.57&   13.29&   15.11\\
SOFT&HII/DUST&   -1.50&   21.46&   14.80&   15.34&   11.35&   12.78&   13.21&   14.11\\
SOFT&solar&   -2.00&   20.99&   14.62&   15.34&   13.20&   13.47&   13.31&   15.10\\
Reddened MF87&solar&-1.3& 22.12& 14.74 &15.34 &12.18 &13.06& 13.26& 15.02\\
\enddata
\tablenotetext{a}{ MF87: \citealt{1987ApJ...323..456M}, AGN: CLOUDY
command, AGN T =1.5e5 k, a(ox) = -1.4, a(uv)=-0.5 a(x)=-1. }
\tablenotetext{b}{Abundance set of H II
region with dust, see  \citet{1991ApJ...374..580B} and CLOUDY HAZY
documentation.} \tablenotetext{c}{Abundance set of Solar, see
\citet{1993PhST...47..133G} and CLOUDY HAZY document.}
\tablenotetext{d}{Total Ni II column density is inferred from
observed column density of Ni II 8395 cm$^{-1}$, using Boltzmann
equation.}
\end{deluxetable}

\section{Origin of the Absorption Gas\label{sec:origin}}
{The physical conditions and location of the absorption gas are well constrained by analyzing the
absorption lines with the aid of photoionization model calculations. It is in the vicinity of
the central engine with a distance of hundreds of parsecs. A laminal geometry of the absorber is inferred
by comparing the physical thickness ($\Delta R \sim 0.02$ \zhy{pc}) with the distance. In addition, the kinematics
of the absorption gas is derived by the profile of isolated lines. The centroid of the absorption lines
observed in SDSS J0802+5513 perches right at the systematic redshift, and the line profile is almost
symmetric in velocity space, spreading from $\sim -750$ km s$^{-1}$ to $\sim +750$ km s$^{-1}$. These
information provides us important clues on the origin of the absorption gas.}

The seemingly relatives to SDSS J0802+5513 are iron low-ionization
broad absorption line (FeLoBAL) quasars, which are defined by the
presence of BALs in excited states of Fe~II and/or Fe~III, in
addition to commonly detected LoBALs, such as Mg~II and Al~III
(e.g. \citealt{2008ApJ...676..857C}).  {The physical conditions of absorbing gas can be well constrained
in a few FeLoBAL quasars, e.g., SDSS J0318$-$0600 \citep{2010ApJ...713...25B,2010ApJ...709..611D}, SDSS J0838+2955 \citep{2009ApJ...706..525M}, FBQS
0840+3633 \citep{2002ApJ...580...54D}, QSO 2359-1241 \citep{2010ApJ...713...25B,2001ApJ...546..140A}. The inferred distance of the absorption gas,
$R$, is typically of sub-kpc or kpc-scale, similar to that of
SDSS J0802+5513. Alike thickness in an order of magnitude of $\Delta R\sim 0.01$
pc is found for these FeLoBAL gas either. Such a laminal geometry, $\Delta R/R \lesssim
10^{-4}$, indicates that the absorption gas is generated right at the place it is observed.
Because, if the absorber were produced in the immediate vicinity of
the central SMBH, it would dissipate long before it arrives at the
inferred location due to Kelvin-Helmholtz instability. Therefore the
absorption gas of SDSS J0802+5513 should be generated in situ as
that of the well-studied FeLoBAL quasars, the \zhy{ionized} gas of which is a
consequence of radiative shocks from interaction of a quasar blast
wave with dense interstellar clumps \citep{2012MNRAS.420.1347F}.}
However, the FeLoBALs are typically blueshifted by several thousand km s$^{-1}$ with
respect to the quasar systematic redshifts, unlike what we observed
in SDSS J0802+5513. {The symmetric line profile is hard to explain by the shock
model, since an impact velocity is always needed to induce the shock
stress.}
{Although most of traditional BALs do not show the redshifted absorbing trough significantly,
\citet{2013MNRAS.434..222H} has identified a small sample of longward-of-system BAL quasars,
which are somewhat resemblant to the absorption troughs in SDSS J0802+5513.} The authors
employed high-velocity infalls or rotationally dominated outflows to
interpret the rarely observed phenomenon. {For individual quasars of the sample (SDSS
J101946.08+051523.7 possibly also SDSS J131637.26+003636.0), the
absorbers are inferred to have a high density ($n_{\rm
e}\approx10^{10.5}$ cm$^{-3}$) and a small distance ($R\le 0.5$
pc) from the the central SMBH, which are very different from that in
SDSS J0802+5513 ($n_{\rm e}\approx10^5$ cm$^{-3}$ and $R \sim
100-250$ pc).} {Therefore, outflowing mechanism cannot explain the line profile
observed in J0802+5513.}

\zhy{Other mechanisms  that may drive the absorption gas in SDSS J0802+5513
involve various stellar processes, such as stellar winds, nova or
supernova (SN) ejecta.} A supernova may cast off gas shells with masses of
$M\gtrsim 1~M_{\odot}$ to velocities of $v\sim 10^3-10^4$ km s$^{-1}$,
\zhy{sufficient to produce} the expansion velocity
of the absorption gas observed in
SDSS J0802+5513. Though the masses of nova shells $M\lesssim
10^{-4}~M_{\odot}$ are much less than that of supernovas, their
expansion velocities are typically $v\sim 10^3$~km~s$^{-1}$
\citep{2006agna.book.....O}, similar to the maximum velocity of the
absorption gas of SDSS J0802+5513. Before exploding as supernovas,
winds of massive stars can remove more than half of the original
mass. To some extreme cases, the terminal velocities of such stellar
winds can reach as high as $v\gtrsim 2\times10^3$~km~s$^{-1}$. The typical
velocities are a few hundred km~s$^{-1}$ \citep{1999isw..book.....L}, \zhy{within} the velocity range of the gas expansion in SDSS J0802+5513.  {Thus the absorption gas of SDSS J0802+5513 could be
generated by the stellar processes in the circumnuclear starburst rings, which are frequently observed in active galaxies.}

An early study of 30 nearby Seyfert galaxies found in their
well-resolved images that $57\%$ have inner rings, $43\%$ have outer
rings, and $\sim 30\%$ have both \citep{1980ApJ...237..404S}. This
finding was confirmed by subsequent UV and optical observations of
Seyfert 2 galaxies (e.g.
\citealt{1998ApJ...505..174G,2001ApJ...546..845G};
\citealt{2001ApJ...558...81C,2004MNRAS.355..273C}). These
observations reveal a typical size of a few hundred pc. About 40\%
nuclear starbursts are very vigorous ($L_{\rm SB}\gtrsim
10^{10}~L_{\odot}$) and compact ($\sim 100$ pc). A similar
conclusion was reached from NIR spectrophotometry of the central
$\sim 300$ pc of 24 Seyfert galaxies \citep{2009MNRAS.400..273R}.
The authors found signatures of young stellar populations in $50\%$
of the Seyfert 2 and most of the Seyfert 1 galaxies. Such
circumnuclear starbursts were suggested to be directly coupled to
the dusty torus, which is the key ingredient of AGN unification
schemes \citep{1993ARA&A..31..473A,1995PASP..107..803U}. The sizes
of circumnuclear starbursts are similar to the estimated distance
of the absorption gas in SDSS J0802+5513 ($R\sim 100-250$~pc from
the galactic center). Interestingly, the gas-to-dust ratio of
$N({\rm H\ I})/E(B-V)=3-9\times 10^{21}~$cm$^{-2}$ mag$^{-1}$, as
the model evaluations in \S4, is close to that of the interstellar
medium (ISM) in the Milky Way ($N({\rm H~I})/E(B-V)=4.8\times
10^{21}~$cm$^{-2}$ mag$^{-1}$, Bohlin et al. 1978). Also the
inferred ratio of log$N({\rm Ca\ II})/E(B-V)\approx 14.0$ for the
absorption gas in SDSS J0802+5513 is within the range of
$13.3\lesssim {\rm log} N({\rm Ca~II})/E(B-V)\lesssim 14.3$ as found in
the Galactic ISM. Thus, we propose that the absorption gas in SDSS
J0802+5513 coexists with the reddening material, which is very
likely the ISM of the host galaxy skirting the dusty torus presumed
by AGN unification models. {This scenario also explains the rareness
of absorption systems like SDSS J0802+5513. The line-of-sight should
rightly penetrate the edge of obscuring material, where the
radiation from central SMBH is mildly reddened.}

{Assuming stars are uniformly distributed in the circumnuclear starburst ring,
we estimate the infalling mass rate induced by stellar process.}
\begin{equation}\label{eq2}
\begin{split}
\dot{M}_{\rm in} &\sim \frac{Mv_{\rm max}/2}{R} \\
&\sim m_{\rm H} \times (N_{\rm H}/2) \times (\frac{2\pi R \times H}{R}) \times (v_{\rm max}/2) \\
&\sim 5~M_{\odot}~{\rm yr}^{-1},
\end{split}
\end{equation}
where $m_{\rm H}=1.67 \times 10^{-24}$~g is the proton mass, $N_{\rm H}\approx
10^{21}-10^{21.5}~{\rm cm}^{-2}$ is the column density, $v_{\rm max}\approx
750$~km~s$^{-1}$ the maximum expansion velocity, $R\approx 100-250$~pc
the distance, and $H\approx R$ the ``height" of the absorption gas.
We estimated $H\approx R$ assuming that the absorption gas were the
periphery of the dusty torus and an absorbed AGN fraction of $f\sim
50\%$\footnote{\citet{2008A&A...490..905H} found that the absorbed AGN fraction
is $f\sim 20-80\%$ for $L_{\rm X}\sim 10^{42}-10^{46}~{\rm erg~s}^{-1}$, and $f$
increases with decreasing X-ray luminosity $L_{\rm X}$}. Intriguingly
enough, such a rough estimate is within the  mass accretion
rate range of SDSS J0802+5513  inferred in \S \ref{sec:emission}. 

\zhy{The proposed stellar processes are closely related to ongoing or recent star formation in the host galaxy.} As a narrow line quasar,  SDSS J0802+5513 may  share some common properties with NLS1s, e.g. enhanced star formation \citep{2010MNRAS.403.1246S}.  In Figure \ref{fig:OII_EW}, we show distribution of
[O II] EWs and [O II]/[Ne III] flux ratios for SDSS DR7 quasars.  \zhy{SDSS J0802+5513 has a larger [O II] EW and a higher [O II]/[Ne III] ratio as compared to the bulk of  DR7 quasars, indicating possible enhanced star formation in this object.}  Under the assumption that all the [O II] emission is contributed by star
formation, we  estimate an upper limit of star formation rate of
11 $M_{\odot}$~yr$^{-1}$ using the calibration of \cite{1998ARA&A..36..189K} \zhy{and tried to distinguish different stellar processes, namely stellar winds and supernova explosions.}

\begin{figure}[!htb]
\hspace*{-1cm}\includegraphics[angle=0,scale=0.6]{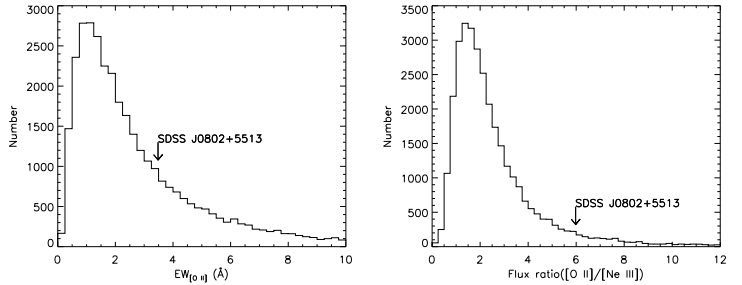}
\caption{\zhy{Distributions of [O II] EW and [O II]/[Ne III] flux ratio for  SDSS DR7 quasars. Values for SDSS J0802+5513 are marked as downward arrows.}
\label{fig:OII_EW} }
\end{figure}
\zhy{Several stellar evolution phases can produce energetic stellar winds.}
Post-main-sequence stars near the ends of their lives
often eject large quantities of mass (10$^{-3}$
$M_{\odot}$~yr$^{-1}$), but the velocities of those wind are typically
10 km s$^{-1}$, far less than the velocity of gas in SDSS J0802+5513. The terminal
velocity of early type stars may meet the velocity requirement, yet with
lower mass loss rate. The wind in early type stars is only efficient at high luminosity
$L>10^4L_{\odot}$ \citep{2000ARA&A..38..613K}  with a mass loss rate of the order
10$^{-5} M_{\odot}$~yr$^{-1}$ or even smaller (10$^{-7}
M_{\odot}$~yr$^{-1}$) for B stars \citep{2014arXiv1401.5511K}. The corresponding stellar mass should
be $>20M_\sun$.  To account for the accretion rate, 5$\times$10$^5$ such stars
are needed. For a Salpeter initial mass function from 0.1 to 125
$M_\sun$ , the required 5$\times$10$^5$ massive stars is equivalent
to a total stellar mass of 10$^8M_\sun$. Assuming a typical lifetime
of 1 Myr for such high mass ($>20M_\sun$) stars, we can deduce that a star formation rate of
100 $M_\sun$ yr$^{-1}$ is needed to supply the mass inflow. The
required star formation rate is approximately an order of magnitude
larger than the observed SFR in SDSS J0802+5513 even assuming all the
[O~II] emission (see Table \ref{tbl:emission}) is contributed by star formation. It seems that the possibility
of stellar winds accretion can be ruled out, although this probably
could be  the case  in AGNs with much lower accretion rate such as Sgr A* \citep{2006MNRAS.366..358C},
where a total stellar mass loss rate of 10$^{-3}$
$M_{\odot}$~yr$^{-1}$ \citep{1997A&A...325..700N}  is sufficient. Either
a large amount of obscured star formation \citep{2012MNRAS.421..486X}
exists given the existence of moderate dust extinction in SDSS
J0802+5513, or mechanisms other than stellar winds should be invoked.

\zhy{On the other hand,} supernova explosions could inject nearly all the mass of their progenitor into their surroundings and has been proposed to work even at
very inner region of AGNs \citep{2010ApJ...719L.148W}. SN explosions observed today are
related to its past star formation history, in the means that a
progenitor with mass $M_*$ will explode $t(M_*)$ years later after its
formation, with $t(M_*)=13(M_*/M_\sun )^{-2.5}$. If a constant star
formation rate of 10 $M_\sun$ yr$^{-1}$ such as in SDSS J0802+5513 holds for
the past 100 Myr, i.e. the typical AGN life cycle \citep{2006ApJ...647L..17W},
all stars with mass larger than 7$M_\sun$ will explode today. Assuming a
Salpeter IMF, a large fraction $f_{\rm c}=0.38$ of star formation mass
will be recycled into the ISM \citep{2010ApJ...719L.148W}. In the case
of SDSS J0802+5513, this adds up to a mass loss rate of 4 $M_\sun$
yr$^{-1}$, given its star formation rate of 10 $M_\sun$ yr$^{-1}$. It
seems that supernova could supply enough gas provided that the star formation in the near past (100 Myr) is at least as intense as what is going on right now in SDSS J0802+5513.
\begin{figure}[!ht]
\includegraphics[scale=0.35,angle=-90]{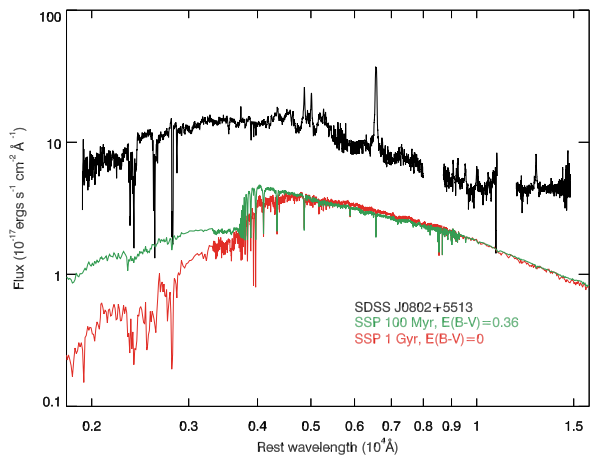}
\caption{Possible stellar populations in SDSS J0802+5513, as constrained by residual flux underneath the saturated absorption troughs of Fe II multiplets UV1, UV2, Mg II doublets and He I  $\lambda$10830.  Overplotted are SSPs from   \citet{2003MNRAS.344.1000B},  normalized to the flux at 10830 \AA\  of SDSS J0802+5513; a 100 Myr SSP reddened by $E(B-V)=0.36$ using SMC-type extinction curve is shown as a green line, while  a 1 Gyr SSP as a red line. \label{fig:ssp}}
\end{figure}

Underlying stellar populations in the \zhy{host galaxies} of AGNs could shed light on their past star formation history.
It is difficult to extract information of the stellar component for high redshift AGNs
as they outshine their hosts. \zhy{For low redshift AGNs, for example in
the quasar He I* absorber IRAS 14026+4341 \citep{2013AJ....145..157J}, \cite{2002ApJ...576...61H} measured the brightness of host galaxy by subtracting the
nucleus in the high-resolution image observed by HST/WFPC2\footnote{Wide Field Planetary Camera 2}}
, yielding
a brightness ratio of host galaxy to nucleus $\sim$12\%.} In the case of SDSS J0802+5513, the
residual fluxes underneath the flat-bottomed absorption troughs are \zhy{at similar levels as observed
in IRAS 14026+4341.} The residual fluxes in multiple wavelengths provide a unique
chance to put constraint on the stellar population of the host
\zhy{under the assumption that most of the residual light can be accounted
for by the stellar component} \citep{1997ApJ...487L.113B}. Four such
troughs are available in SDSS J0802+5513, namely Fe II UV1 and UV2 at
2600 \AA\ and 2400 \AA, respectively, Mg II doublet at \zhy{around 2800 \AA\ } and He I* $\lambda$10830, spanning
from NUV to NIR. In \citet{2000AJ....120.2859N}, four low-ionization BAL troughs in a reddened
BAL quasar F1556+3517, located at 1860--2800 \AA\, are utilized to put
constraint on stellar populations. From the observed upper limit on
the strength of the 4000 \AA\ break, the author favors a reddened 50 Myr stellar
population over a 1 Gyr one. Besides the evidence proposed in \citet{2000AJ....120.2859N}, further
constraint can be put on SDSS J0802+5513 with the aid
of our new \zhy{NIR} spectroscopy of He I* $\lambda$10380. As shown in Figure
\ref{fig:ssp}, SSPs with ages of 100 Myr and 1 Gyr from \citet{2003MNRAS.344.1000B} are normalized to the
flux level at the bottom of He I* $\lambda$10830 trough. \zhy{Although}
older SSPs \zhy{can} mimic reddened young SSPs in
optical--NIR band (i.e. the long-known age-dust degeneracy), the UV flux of older SSPs 
\zhy{drops} rapidly and fail to fit the UV absorption troughs in SDSS J0802+5513.
Using saturated absorption lines as coronagraphs, \zhy{we  inferred
that the host galaxy should consist of a significant population of reddened young
(several hundred Myr or less) stars, indicating the recent star formation is probably as intense as required for supernova explosion powered winds.}
The above discussion should be understood with the caveat that, without further observations,
we \zhy{cannot} tell if some of the light underneath the absorption trough is contributed  by scattered light of the background quasar.

We conclude that supernova explosion seems to be most promising paradigm
that drives the gas flow in SDSS J0802+5513 and we will end our
discussion with a possible destiny of the gas following the line of reasoning. While
the inflowing gas in SDSS J0802+5513 keeps feeding its central black
hole, the outflowing gas would reach the outskirt out its host in
50 Myr, when the host will evolve into a post-starburst galaxies
given the estimated age of underlying stellar population. The large
scale outflows are ubiquitous in post-starbursts and the origin of
these outflows remains a myth, both AGNs and starbursts have been proposed to
driven the wind. Study on objects similar to  SDSS J0802+5513 could help to
resolve the problem.
\section{Summary and Future Perspectives\label{sec:final}}
We present detailed analysis of SDSS spectrum and newly obtained MMT, P200 and LBT spectra for SDSS J0802+5513. The object is classified
as a NLS1 based on the widths of Balmer emission lines and  Fe II
emission strength. It's moderately dust reddened with an extinction
of $E(B-V)=0.36$. Its spectra show rare absorption lines of He I*
and Fe II*, as well as lines from Ca II, Mg I, Mg II, Ni II and Mn
II. The absorption lines show identical profile ranging from -750 to
750 km s$^{-1}$, with its centroid at the same redshift as that determined from
emission lines. This object is the \zhy{first unshifted He I* min-BAL reported.} Extensive
photoionization models are calculated using CLOUDY, and He I* is shown
to be a good indicator of ionization parameter. With the aid of the models, the physical parameters of the
absorption gas are constrained to be log$U\sim-1.8$, $N_{\rm H}\sim10^{21}$
cm$^{-2}$, $n_{\rm H}\sim10^5$ cm$^{-3}$. The gas is estimated to be
$R\sim100$--250 pc away from central black hole.  Various origins of the absorption gas are discussed and current observations suggest stellar processes are at work in driving the gas flow. Without further
knowledge whether there exists hidden star formation in SDSS J0802+5513, mass losses from stellar winds of high mass stars  fail to account for the deduced mass inflowing rate of 5 $M_{\odot}$~yr$^{-1}$. The current
data support supernova explosions as the main paradigm funneling the
gas in SDSS J0802+5513.
\begin{figure*}[!ht]
\hspace*{-1cm}\vspace*{0cm}
\includegraphics[angle=0,scale=1]{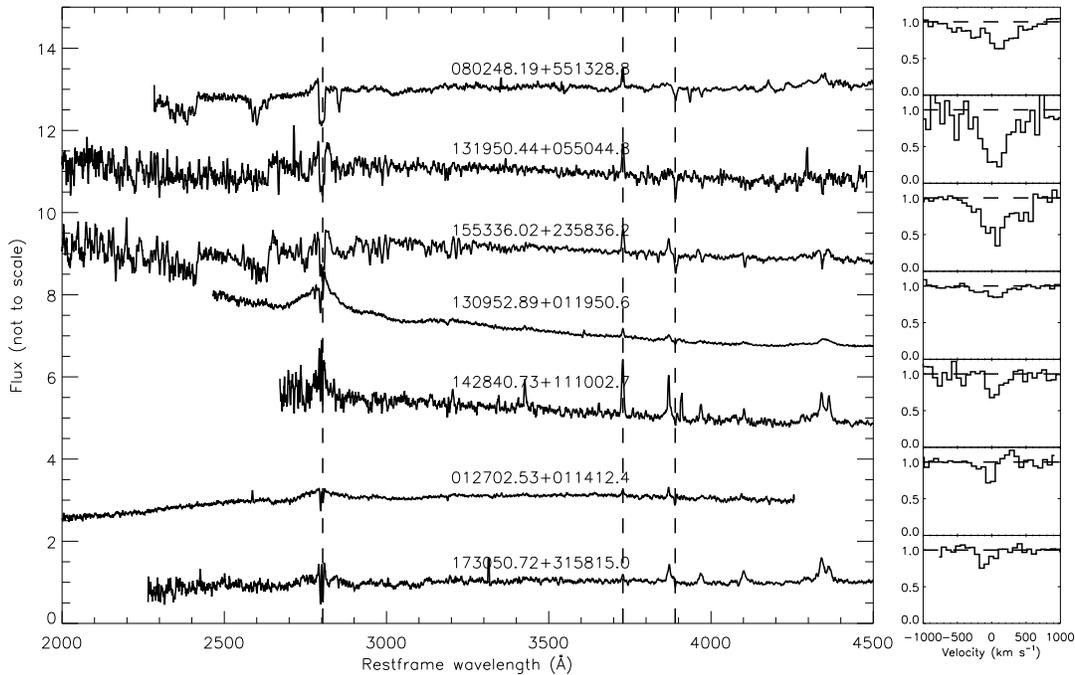}
\caption{\zhy{SDSS Spectra of unshifted He I* absorption line quasars. Normalized spectra of He I*$\lambda$3889 are shown on the right.\label{fig:heIcomposite}}}
\end{figure*}
Follow-up observations of SDSS J0802+5513 are needed to better study  its
properties: (1) mid-IR spectroscopies are needed to confirm if there is
any hidden star formation, which will put firm constraint on the
driving source of the gas flow; (2) polarization observations are needed
to see if any polarized (scattered) light exists underneath the absorption trough
and narrow band imaging centered at the absorption troughs will help to
constrain the stellar population; (3) High resolution spectroscopies
are useful to resolve the heavily blended absorption lines and to improve column density measurements;  and 4) X-ray observations are helpful to determine total column density of the absorption gas.


\zhy{Finally, we note that though very rare, SDSS J0802+5513 has similars. We show the spectra of  the 7 unshifted He I* absorber candidates in Figure \ref{fig:heIcomposite}. Interestingly, all but one (SDSS J130952.89+011950.6) show a red color, like SDSS J0802+5513. Given the ubiquity of star formation ring in low redshift AGNs (Simkin et al. 1980), we might expect to detect much more similar objects than observed.  The rarity could have an important implication on the geometry of the absorbers. If the nuclear star formation is co-plane with the dusty torus of quasars, optical spectroscopic surveys like SDSS will have little change to detect such dusty absorbers. Only when our sight lines are coincidently penetrating the  edge of
the torus with a moderate optical depth, such as those partially
obscured quasars \citep{2005ApJ...620..629D}, can we detect absorbers like SDSS J0802+5513.  Studies on these objects could reveal  (1) the true nature of quasars with unshifted He I* absorption lines,  (2) their possible connection with partially obscured quasars  and (3) the role that starburst driven gas flows play in the coevolution of SMBH and stellar bulges.}

\acknowledgments This work is supported by the SOC program (CHINARE2012-02-03), the NSFC grant (NSF11033007), National Basic Research Program of China (973 Program, 2013CB834905, 2015CB857005) and  Fundamental Research Funds for the Central Universities (WK 2030220010). P.J. acknowledges supports from the National Science Foundation of China with Grants NSFC 11233002 and NSFC11203022, the Fundamental Research Funds for the Central Universities and the China Postdoctoral Science Foundation. Data from NED,
NIST and SDSS were used. This research also uses data obtained
through the Telescope Access Program (TAP), which is funded by the
National Astronomical Observatories, Chinese Academy of Sciences,
and the Special Fund for Astronomy from the Ministry of Finance.
Observations obtained with the Hale Telescope at Palomar Observatory
were obtained as part of an agreement between the National
Astronomical Observatories, Chinese Academy of Sciences, and the
California Institute of Technology.


\bibliography{./all_no_abstract}
\end{document}